\documentclass[nofootinbib,aps,10pt,twocolumn,superscriptaddress]{revtex4}
\usepackage[dvips]{graphicx}

\usepackage{array,amsmath,amsthm,feynmf}
\usepackage{bm}
\usepackage{times}
\usepackage{epsfig}
\usepackage{graphicx}
\usepackage{color}
\usepackage{cancel}
\usepackage{amssymb}
\usepackage{textcomp}

\def\OMIT#1{}

\newcommand{\bea}{\begin{eqnarray}}
\newcommand{\eea}{\end{eqnarray}}

\newcommand{\be}{\begin{equation}}
\newcommand{\ee}{\end{equation}}

\DeclareMathOperator{\Tr}{Tr}

\newcommand{\veff}{V_\mathrm{eff}}
\definecolor{dRed}{rgb}{.7,0,0}

\newtheorem{theorem}{Theorem}

\usepackage{hyperref}

\hypersetup{
   colorlinks=true,       % false: boxed links; true: colored links
   linkcolor=black,          % color of internal links
   citecolor=black,        % color of links to bibliography
   filecolor=black,      % color of file links
   urlcolor=black           % color of external links
}

\begin{document}
\smash{\hspace{6 cm} NPAC-11-01}\vspace{-5cm}
\title{\Large Baryon Washout, Electroweak Phase Transition, and Perturbation Theory}
\vspace{4.0cm}
\author{Hiren H. Patel}
\email{hhpatel@wisc.edu}
\affiliation{
{University of Wisconsin-Madison, Department of Physics} \\
{1150 University Avenue, Madison, WI 53706, USA}}
\author{Michael J. Ramsey-Musolf}
\email{mjrm@physics.wisc.edu}
\affiliation{
{University of Wisconsin-Madison, Department of Physics} \\
{1150 University Avenue, Madison, WI 53706, USA}}
\affiliation{
{Kellogg Radiation Laboratory, California Institute of Technology}\\
{Pasadena, CA 91125 USA}}
\date{\today}
%%%%%%%%%%%%%%%%%%%%%%%%%%%%%%%%%%%%%%%%%%%%%%%%%%%%%%%%%%%%%%%%%%%%
\begin{abstract}
We analyze the conventional perturbative treatment of sphaleron-induced baryon number washout relevant for electroweak baryogenesis and show that it is not gauge-independent due to the failure of consistently implementing the Nielsen identities order-by-order in perturbation theory.  We provide a gauge-independent criterion for baryon number preservation in place of the conventional (gauge-dependent) criterion needed for successful electroweak baryogenesis.  We also review the arguments leading to the preservation criterion and analyze several sources of theoretical uncertainties in obtaining a numerical bound.  In various beyond the standard model scenarios, a realistic perturbative treatment will likely require knowledge of the complete two-loop finite temperature effective potential and the one-loop sphaleron rate.
\end{abstract}
%%%%%%%%%%%%%%%%%%%%%%%%%%%%%%%%%%%%%%%%%%%%%%%%%%%%%%%%%%%%%%%%%%%%
\maketitle
%%%%%%%%%%%%%%%%%%%%%%%%%%%%%%%%%%%%%%%%%%%%%%%%%%%%%%%%%%%%%%%%%%%%%
%\tableofcontents
\section{Introduction}

Explaining the cosmic baryon asymmetry remains an open problem at the interface of cosmology with particle and nuclear physics.  Among the several proposed scenarios, electroweak baryogenesis (EWB) is particularly interesting as it is conducive to experimental tests.  The mechanism requires a strong first order electroweak phase transition (EWPT) that proceeds via bubble nucleation at temperatures $\sim$100 GeV and sufficient CP violation. CP-violating interactions at the bubble walls induce a net density of left handed fermions ($n_L$) that biases electroweak sphalerons into generation of baryon number density. As the transition proceeds,  baryon number diffuses into the interiors of the expanding bubbles where electroweak symmetry breaking slows the sphaleron transitions.  The latter leads to baryon number erasure as the sphalerons try to restore chemical equilibrium.  Thus,  successful EWB requires that the sphaleron rate is sufficiently suppressed inside the bubbles to prevent this washout. 

In the standard model (SM), the effects of CP-violation associated with the Cabibbo-Kobayashi-Maskawa (CKM) matrix are too feeble to generate a sufficiently large $n_L$ to bias the sphalerons in the first place. Even if this were not the case, however, the finite temperature dynamics of the SM scalar sector do not allow for a first order EWPT as needed to prevent erasure of an initial baryon asymmetry. Consequently, successful EWB requires both additional sources of electroweak CP-violation and an augmented scalar sector. Experimentally, 
searches for the permanent electric dipole moments of the neutron, electron, and neutral atoms provide a probe of the requisite new CP violation, while searches for new scalar particles at the CERN Large Hadron Collider may reveal the ingredients needed for the strong first order EWPT that would preserve the resulting baryon asymmetry.

In this work we focus on the theoretical analysis of the washout of the baryon asymmetry during the phase transition.  The most theoretically robust approach to the study of EWPT dynamics involves Monte Carlo lattice simulations. Several such studies have been carried out in the SM, yielding results for the critical temperature, bubble nucleation rate, and sphaleron rate \cite{Fradkin:1978dv,Ambjorn:1990wn,Kajantie:1996qd,Kajantie:1996mn,Csikor:1998eu,Aoki:1999fi,Moore:1998swa}. These quantities depend critically on the Higgs quartic self coupling, which also governs the value of the Higgs boson mass, $m_H$. Consequently, there exists a tight connection between $m_H$ and baryon number washout. Lattice results indicate that a strong first order EWPT as needed for baryon number preservation requires $m_H\lesssim 70$ GeV, well below the current LEP lower bound of $114.4$ GeV\cite{Barate:2003sz}. Thus, EWB is only viable in the presence of an extended scalar sector that is not subject to the current LEP limits. 

A number of extended scalar sector scenarios have been analyzed that may successfully remedy the SM shortcomings. The most widely considered are supersymmetric extensions (see {\em e.g.} \cite{Quiros:1999jp,Huber:2001xf,Carena:2008vj,Kang:2004pp,Chung:2010cd,Funakubo:2009eg,Chiang:2009fs} and references therein), though recent work has also included analysis of non-supersymmetric models (see {\em e.g.} \cite{Profumo:2007wc,Ham:2005ej,Ham:2010ha} and references therein) . While there exist a handful of Monte Carlo studies of phase transition dynamics in the minimal supersymmetric standard model (MSSM) \cite{Laine:1998qk,Csikor:2000sq}, the majority of phase transition studies in extended scalar sector scenarios have relied on the use of perturbation theory.  Moreover, non-perturbative calculations in the SM and MSSM have generally concentrated on the thermodynamic properties of a possible EWPT (such as the critical temperature) rather than directly computing the sphaleron rate that governs the degree of baryon number washout (for important exceptions, see \cite{Ambjorn:1990wn,Moore:1998swa,Moore:2010jd}). 

This situation is not surprising, given the numerical cost in performing non-perturbative computations. At present, it is simply not feasible to survey the broad range of extended scalar sector models and to analyze in each the often large, multidimensional parameter space  using Monte Carlo methods. In order to identify the EWPT-viable parameter space regions and identify their phenomenological signatures for collider searches, the reliance on perturbation theory appears to be unavoidable.  Thus, it is desirable to arrive at the most theoretically sound perturbative treatment of baryon number erasure in order to identify candidate scenarios that are most promising for EWB and to determine their prospective experimental signatures.

In what follows, we revisit the conventional, perturbative treatment of baryon number washout and observe that it suffers from a serious theoretical shortcoming, namely, dependence on the choice of gauge. Consequently, there exists reason to question conclusions drawn about the EWB-viability of scenarios analyzed according to the conventional perturbative treatments. We subsequently show that there exist means of arriving at a gauge-independent, perturbative treatment of baryon number washout and we analyze one such approach in detail. We argue that the use of this approach can provide a theoretically unambiguous indication of the conditions under which a given scenario may lead to baryon number preservation and can point to the relevant, but more limited, regions of parameter space that should be explored with Monte Carlo computations. In the course of our discussion, we also revisit the approximate criterion used to evaluate the efficacy of baryon number preservation---henceforth referred to as the ``baryon number preservation criterion'', or BNPC.  We also argue that a realistic BNPC may be more stringent than the one widely used in recent years and that there exist important theoretical uncertainties that call for future study. As a practical matter, we provide an approximate BNPC that differs from the standard formula used in the literature and that takes into account the foregoing issues.

\subsection{Brief Overview}
\label{sub:over}

Before analyzing these considerations in detail, we provide a brief overview. If a first order EWPT occurs, it commences via the formation of broken electroweak symmetry bubbles at the nucleation temperature $T_N$.  The latter typically falls just below the critical temperature $T_C$ at which the broken and unbroken minima of the scalar effective potential become degenerate.  Inside the  broken symmetry phase, the depletion of baryon number density $n_B$ is expected to follow a first order rate law
\begin{gather}\label{eq:ratelaw}
\frac{\partial n_B}{\partial t}=-k(T) n_B\,,
\end{gather}
where $k(T)$ is the temperature-dependent rate constant\cite{Kramers:1940zz}
\begin{equation}\label{eq:rateConst}
k(T)=-\frac{13 n_f}{2} \frac{\Gamma_\text{sph}(T)}{VT^3}\sim A(T) e^{-\Delta E_\text{sph}/T}\,,
\end{equation}
governed by the temperature $T$, number of fermion generations $n_f$, and most importantly the sphaleron rate per unit volume $\Gamma_\mathrm{sph}/V$ that depends exponentially on the sphaleron energy relative to that of the electroweak vacuum, $\Delta E_\text{sph}$. Additional, non-exponential dependence on $T$ is contained in the prefactor $A(T)$.

The extent to which an initial baryon asymmetry is erased over the duration of the transition, $\Delta t_\text{EW}$, is characterized by the \lq\lq washout factor"
\begin{equation}
\label{eq:washoutfactor}
S=\frac{n_B(\Delta t_\text{EW})}{n_B(0)}\,.
\end{equation}
Assuming for illustrative purposes that $k(T)$ is approximately constant during the transition, integration of (\ref{eq:ratelaw}) yields
\begin{equation}
\ln S\sim e^{-\Delta E_\text{sph}/T_C}
\end{equation} 
where we have taken the critical temperature $T_C$ as a characteristic temperature of the transition.  
The magnitude of an initial baryon asymmetry, determined by the strength of the CP-violating interactions and transport dynamics, will dictate a lower bound on $S$ 
\begin{equation}
S> e^{-X}\,,
\end{equation}
below which the final baryon asymmetry would be smaller than the observed value. Baryon number preservation thus implies roughly a requirement on $\Delta E_\text{sph}/T_C$ and thereby on the underlying particle physics that determines the critical temperature and the sphaleron energy at that temperature.

In the conventional perturbative analysis, the sphaleron energy $\Delta E_\text{sph}$ is expressed in terms of $\phi_\text{min}(T_C)$,  the classical field that minimizes the finite temperature scalar effective potential at the critical temperature. Taking $X\approx 10$ (which we question below) and including the dependence on the $\Delta t_\mathrm{EW}$ and other factors that enter the rate equation via (\ref{eq:ratelaw}) (see Appendix \ref{sec:washout} for details), one arrives at the following requirement:
\begin{equation}\label{eq:exp}
\frac{\phi_\text{min}(T_C)}{T_C}\gtrsim1.0\, \qquad \text{(conventional criterion)}\,.
\end{equation}
As we discuss below, this criterion is open to question for three reasons:
\begin{itemize}
\item[(1)] The classical field $\phi_\text{min}(T)$ is inherently gauge-dependent at any temperature.
\item[(2)] The method of computing $T_C$ introduces additional gauge dependence that does not compensate the gauge-dependence of $\phi_\text{min}(T_C)$.
\item[(3)] The choice of $X\approx 10$ is overly optimistic for most scenarios, and there exist additional uncertainties associated with computing the prefactor $A(T)$ and integrating (\ref{eq:ratelaw}).
\end{itemize}

In what follows, we show that it is possible to obtain a gauge-independent perturbative estimate of the sphaleron rate and critical temperature and review both the theoretical uncertainties alluded to in point (3) as well as the choice of $X$. In brief,
\begin{itemize}
\item[(1)] The sphaleron rate $\Gamma_\text{sph}/V$ is obtained by evaluating the temperature-dependent effective action of the sphaleron.  A gauge-independent, perturbative computation of this quantity can be performed using a dimensionally reduced 3D effective theory where only the gauge-independent $\mathcal{O}(T^2)$ terms are included in the effective action.  As a result, the theory contains a gauge-independent energy scale $\bar{v}(T)$ that is non-vanishing below a temperature $T_0$ (different from $T_C$) and that characterizes the sphaleron energy.  

Although additional gauge-independent, one-loop thermal contributions to $\Gamma_\text{sph}/V$ 
have been obtained numerically  for the minimal SM \cite{Carson:1990jm,Baacke:1993aj,Baacke:1994ix}, similar perturbative computations do not exist for its extensions. Consequently, the conventional practice has been to estimate these contributions by replacing the gauge-independent sphaleron scale $\bar{v}(T)$ with the gauge-dependent Higgs field $\phi_\text{min}(T)$ that minimizes the full gauge-dependent effective potential $V_\text{eff}(\phi, T)$.  As a result, the standard practice leads to a gauge-dependent estimation of the sphaleron rate regardless of the temperature. Clearly, a gauge-independent calculation requires utilizing the gauge-independent scale ${\bar v}(T)$ rather than $\phi_\text{min}(T)$.

\item[(2)] The critical temperature is obtained from the behavior of the finite-temperature effective potential $V_\text{eff}(\phi,T)$.  Bubble nucleation characteristic of a first order phase transition requires the existence of a barrier separating the electroweak symmetry-broken and -unbroken minima in $\veff(\phi,T)$. At one-loop order in the SM, this barrier is generated by non-analytic $\propto T\phi^3$ terms in the potential. Since the early work of Dolan and Jackiw, however, it has been known that this non-analytic part of the potential is gauge-dependent \cite{Dolan:1973qd}, and the early papers argued that one should drop these terms from $\veff(\phi,T)$ when deriving $T_C$. Since doing so allows only for a second order phase transition, the common practice in EWB studies has been to retain the non-analytic terms in $V_\text{eff}(\phi,T)$.  This causes $T_C$ to inherit the gauge-dependence.

A gauge-independent value for $T_C$ can be obtained in perturbation theory by either using a gauge-invariant source term $j \Phi^\dag\Phi$ in the generating functional or by working with a source term $j \Phi$ which not gauge-invariant, and consistently implementing the Nielsen's identity\cite{Nielsen:1975fs}. In this work, we follow the latter approach and defer an exploration of the former to subsequent study. Nielsen's identity, derived shortly after the work of Dolan and Jackiw, implies that  the exact effective action $\Gamma_\text{eff}$ and potential $V_\text{eff}$ are gauge-independent at their extremal points, but that mean fields ($\phi_\text{min}$) derived from them remain gauge-dependent.  Since $T_C$ is defined in terms of minima of $V_\text{eff}$ and the sphaleron rate is derived from the saddle point of $\Gamma_\text{eff}$, both quantities are expected to be gauge-independent.  However, na\"ive truncation of the perturbative series introduces spurious gauge-dependence in $T_C$. As we show below, it is possible to avoid this spurious gauge-dependence  by determining $T_C$  from the extrema of $V_\text{eff}(\phi,T)$ in a manner consistent with the $\hbar$-expansion.

\item[(3)] Provided that one obtains a gauge-independent computation of the sphaleron rate and critical temperature, there exist remaining uncertainties in using these quantities to derive a BNPC. For example, the choice of $X\sim 10$ derives from early work by Shaposhnikov\cite{Shaposhnikov:1987tw}, who estimated the SM baryon asymmetry from the equation of motion for Chern-Simons number. The computation required an estimate of the quark-antiquark asymmetry $\delta_\mathrm{ms}$---produced by CP-violation associated with the CKM matrix---that enters as a source term in the Chern-Simons equation of motion. While a range of possible values for $\delta_\mathrm{ms}$ were considered in Ref.~\cite{Shaposhnikov:1987tw}, the most generous possibility allows for a significant overproduction of baryons in the SM and requires a large washout of the baryon asymmetry ($X\sim 10$). It was subsequently argued, however, that CP-violation from the CKM matrix would have been far too small to overcome even a washout factor of $\mathcal{O}(1)$ (for a discussion, see {\em e.g.} Ref.~\cite{Dine:1990fj}). 

Recent attention has focused on the production of CP-violating asymmetries in supersymmetric models, generated by the scattering of superpartners from the bubble walls (see Ref.~\cite{Chung:2009qs} and references therein). Constraints on the supersymmetric CP-violating phases imposed by EDM searches generally imply that these asymmetries are right on the edge, at best, of inducing the observed baryon asymmetry. Consequently, in these scenarios, a washout factor closer to $\mathcal{O}(1)$ (or $X \ll 1$) is likely to be more realistic. In addition, the work of Ref.~ \cite{Moore:1998swa} suggests that at least in the Standard Model, the duration of the transition is likely to be closer to $\Delta t_\mathrm{EW}\sim 10^{-3} t_H$, where $t_H$ is the Hubble time, rather than $\Delta t_\mathrm{EW}\sim t_H$ as assumed in arriving at the criterion in Eq.~(\ref{eq:exp}). 

There exist additional uncertainties associated with the frequency of the unstable sphaleron mode (known from the work of Ref.~\cite{Carson:1990jm}  but not generally included) and the evaluation of the sphaleron fluctuation determinant (known and generally considered) that should be taken into account when seeking to determine whether or not the underlying particle physics dynamics lead to sufficient preservation of an initial baryon asymmetry. The approximate BNPC that we derive below attempts to take these considerations fully into account.
\end{itemize}

The primary implications of our analysis of these issues is to replace the gauge-dependent BNPC (\ref{eq:exp}) with the gauge-independent one and to include the aforementioned theoretical uncertainties. In particular, in the LHS one should make the replacement
\begin{equation}
\label{eq:replace}
\frac{\phi_\text{min}(T_C;\,\xi)}{T_C(\xi)} \enspace\,\longrightarrow\enspace\, \frac{\bar v(T_C)}{T_C}\ \ \ ,
\end{equation}
where $\xi$ is the gauge-fixing parameter and where the quantities to the right of the arrow are $\xi$-independent. On the RHS of the BNPC (\ref{eq:exp}), one replaces unity by a range of bounds determined by an appropriate choice of $X$ and the other theoretical inputs.  The former will depend on the magnitude of the initial baryon number density $n_B(0)$ generated by CP-violating interactions at the bubble walls as well as particle number changing reactions that extend into the unbroken phase where sphaleron transitions are unsuppressed. We will provide an expression for the BNPC that takes these considerations into account in Section \ref{sec:BNPC}.

In organizing our discussion of these issues in the remainder of the paper, we will be somewhat pedagogical since the issue of gauge-independence, though clearly of fundamental importance, is subtle. Consequently, some of the material consists of a review of earlier work, where we collect and recast in this context results that have been previously derived in the literature. In addition, we  observe that although the complete evaluation of  $\Gamma_\text{sph}$ is vastly more difficult than determining  $T_C$, it appears that at least in the SM  the numerical impact of the treatment of $T_C$ appears to be significantly more important than the one-loop contributions
to $\Gamma_\text{sph}$ associated with fluctuations about the sphaleron.  Consequently, most of our discussion is focused on $T_C$ and $V_\text{eff}$.

To that end, in section \ref{sec:GaugeProblem}, we discuss the effective potential and the Nielsen identities using a general model that accommodates an arbitrary extension of the SM.  We identify the gauge-independent quantities derivable from the potential.  We also discuss how a failure to consistently implement the Nielsen identities leads to gauge-dependence in $T_C$.  We subsequently outline our gauge-independent treatment of the critical temperature in section \ref{GIanalysisSection}, and discuss the gauge-independent sphaleron rate in section \ref{sec:sphaleron}.  We provide a concrete, numerical application of the gauge-independent analysis of the SM in section \ref{sec:SM}, and we discuss the practical implications and theoretical limitations of our procedure -- along with the modified BNPC --  in section \ref{sec:BNPC}.  In that section, we also examine the dependence of the BNPC on a variety of theoretical inputs. Finally, we conclude in section \ref{sec:conclusions}.  Several formal results are contained in the Appendix.

\section{Gauge Problem in the Standard Analysis}\label{sec:GaugeProblem}
We begin our discussion by focusing on the effective potential $V_\text{eff}$, since it from this quantity that the critical temperature $T_C$ and the background field $\phi_\text{min}$ are derived.  The standard practice has been to work in Landau gauge ($\xi=0$), but we will leave $\xi$ arbitrary to illustrate how gauge dependence of $T_C$ and $\phi_\text{min}$ appear in the standard analysis. For pedagogical purposes, we will first derive the zero-temperature effective potential in a general gauge and follow up with its generalization to finite temperature.  After briefly discussing implications of the gauge problem in the standard analysis, we will discuss how Nielsen's identities  shed light on this problem.

\subsection{Zero temperature effective potential}\label{SectionZeroT}
We will work in a very general 4D theory\cite{Weinberg:1973ua} (\lq\lq general model") containing an arbitrary number, $n$, of real scalar degrees of freedom, which are assembled into a column vector, $\Phi_i(x)$ (with $i=1,\ldots,n$), transforming under a general (possibly reducible) real representation of an arbitrary (semi-simple) gauge group, $\mathcal{G}$ with $N$ generators. In the real representation, the generators $t^a$, $a=1,\ldots,N$ are purely imaginary and antisymmetric.  For convenience a negative imaginary unit, $-i$, is factored out, so that generators are purely real and antisymmetric $t^a=-i\,T^a$.  All gauge bosons of the theory are assembled into a vector, $A_\mu^a$, so that the generating functional and Lagrangian of the theory may be written in a  compact manner:
\begin{gather}
\label{genFunctional}
Z[j]=\int\mathcal{D}\Phi\,\mathcal{D}A \,e^{i\int d^4x\, \mathcal{L}(x;\,j)}\\
\label{VGM}\mathcal{L}=\frac{1}{2}D_\mu\Phi_i D^\mu\Phi_i-\frac{1}{4}F_{\mu\nu}^a F^{\mu\nu\, a}-V(\Phi)+j_i\Phi_i
\end{gather}
Our convention for the gauge covariant derivative and field strength tensor is\footnote{In general, different gauge coupling constants are assigned to each simple subgroup of $\mathcal{G}$.  To avoid notational clutter, all coupling constants are identically denoted as $g$ and it should be clear which one we refer to, based on the generator or gauge field appearing adjacent to it.} $D_\mu\Phi_i = \partial_\mu\Phi_i+A_\mu^a (gT^a)_{ij}\Phi_j$, and $F_{\mu\nu}^a=\partial_\mu A_\nu^a-\partial_\nu A_\mu^a-g f^{abc}A_\mu^b A_\nu^c$, where $f^{abc}$ are the structure constants of $\mathcal{G}$.   The effective potential is obtained by passing from the generating functional of connected Greens functions $W[j]=-i\ln Z[j]$ to the generating functional of proper (1PI) vertices, $\Gamma[\phi_\text{cl}]$ via a Legendre transformation 
\begin{equation}
\label{effAction} \Gamma[\phi_\text{cl}(x)]=W[j]-\int d^4x\, j(x)\phi_\text{cl}(x)\,,
\end{equation}
where $\phi_\text{cl}(x)\equiv\partial W[j]/\partial j(x)$ is the classical field conjugate to the source, $j(x)$.  For space-time homogenous classical fields $\phi_\text{cl}(x)\equiv\phi_\text{cl}$, we obtain the effective potential, $\Gamma(\phi_\text{cl})=-(\text{vol})\,V_\text{eff}(\phi_\text{cl})$. To properly define the functional integral in (\ref{genFunctional}), we must factor out a set of redundant gauge configurations. Doing so introduces the gauge-dependence that is the central focus of our analysis. 

To make this gauge-dependence explicit, we derive the  effective action following the background field method, where the scalar fields are (homogeneously) shifted as a change of variables under the functional integral in (\ref{genFunctional}).  We write $\Phi_i(x)=\bar\phi_i+\phi_i(x)$,
where $\bar\phi_i$ are space-time independent background fields\footnote{More precisely, one should write the background field as being dependent on the source, $\bar\phi\equiv\bar\phi(j)$.  The dependence on $j$ is chosen so that, upon performing the Legendre transformation, $\bar\phi$ precisely coincides with $\phi_\text{cl}$, facilitating the evaluation of the perturbative loop expansion of the effective action $\Gamma[\phi_\text{cl}]$ \cite{Jackiw:1974cv}.} and $\phi_i(x)$ represent quantum fluctuations around the background field.  

Organized by powers of quantum fields, the resulting lagrangian after shifting  and before gauge-fixing is
\begin{multline}\label{homBGlag}
\mathcal{L}(x;j)=-V(\bar\phi)+j\bar\phi+\phi_i\Big(-\frac{\partial V}{\partial \Phi_i}\Big|_{\bar\phi}+j_i\Big)\\
+\frac{1}{2}\phi_i\big[-\partial^2-M_{ij}^2(\bar\phi)\big]\phi_j-\partial^\mu\phi_i A^a_\mu(gT^a\bar\phi)_i\\
+\frac{1}{2}A_\mu^a\big[(\partial^2g^{\mu\nu}-\partial^\mu\partial^\nu)\delta^{ab}+m_A^2(\bar\phi)^{ab}g^{\mu\nu}\big]A_\nu^b,
\end{multline}
where 
\begin{align*}
M_{ij}^2(\bar\phi)&=\partial^2 V/\partial\phi_i\,\partial\phi_j|_{\bar\phi}\\
m_A^2(\bar\phi)^{ab}&= (g{  T}^a\bar\phi)_i(g{  T}^b\bar\phi)_i
\end{align*}
are field-dependent mass matrices for the scalar and vector gauge bosons, respectively.  Terms cubic and higher order in quantum fields are not shown.
 
We now impose the background-field dependent gauge condition\footnote{It is more general to choose the fixed-vector gauge condition: 
$ \partial_\mu A^{\mu\, a}-\xi\phi_i v_i^a=0$ originally proposed in \cite{Dolan:1974gu}.  However, it was pointed out in \cite{Fukuda:1975di} that the fixing condition does not allow for a consistent analysis of the vacuum structure under the homogeneity condition.}, 
\begin{equation}\label{eq:gfCondition}
\mathcal{F}^a\equiv \partial_\mu A^{\mu\, a}-\xi\phi_i (g{T}^a\bar\phi)_i=0 
\end{equation}
on the quantum fields. This choice, motivated by the $R_\xi$ gauges, has the advantage that adding $\mathcal{L_\text{gf}}=-(\mathcal{F}^a)^2/2\xi$
to the Lagrangian cancels the $-\partial^\mu\phi_i A^a_\mu(g{\tilde T}^a\bar\phi)_i$ mixing term appearing in (\ref{homBGlag}).  The compensating ghost Lagrangian is
\begin{multline}\label{homFPghost}
\mathcal{L}_\text{gh}=\eta^{\dag a}\big[-\partial^2\delta^{ab}-\xi m_A^2(\bar\phi)^{ab}\big] \eta^b\\
+gf^{abc}(\partial^\mu\eta^{\dag a})\eta^b A_\mu^c-\xi(g{T}^a\bar\phi)_i\eta^{\dag a}\eta^b (g {T}^b_{ij} \phi_j)\,.
\end{multline}

Upon adding $\mathcal{L}_\text{gf}$ and $\mathcal{L}_\text{gh}$ to (\ref{homBGlag}), the gauge-fixed generating functional is
\begin{equation}
Z[j]=\int \mathcal{D}\phi\,\mathcal{D} A\, \mathcal{D}\eta\, \mathcal{D}\eta^\dag e^{i\int d^dx\big(\mathcal{L}(x;j,\xi)\big)}\,,
\end{equation}
with
\begin{multline}\label{gfLag}
\mathcal{L}(x;\xi,j)=-V(\bar\phi)+j\bar\phi+\phi_i\Big(-\frac{\partial V}{\partial \Phi_i}\Big|_{\bar\phi}+j_i\Big)\\
\shoveleft+\frac{1}{2}\phi_i\big[-\partial^2-M_{ij}^2(\bar\phi)-\xi m_A^2(\bar\phi)_{ij}\big]\phi_j\\
\shoveleft+\frac{1}{2}A_\mu^a\big[\big(\partial^2g^{\mu\nu}-(1-\frac{1}{\xi})\partial^\mu\partial^\nu\big)\delta^{ab}+m_A^2(\bar\phi)^{ab}g^{\mu\nu}\big]A_\nu^b\\
+\eta^{\dag a}\big[-\partial^2\delta^{ab}-\xi m_A^2(\bar\phi)^{ab}\big]\eta^b+\ldots\,,
\end{multline}
where we have suggestively written 
\be
\label{eq:mAij}
\xi m_A^2(\bar\phi)_{ij}=\xi (gT^a\bar\phi)_i(gT^a\bar\phi)_j
\ee
as the additional gauge-dependent scalar boson mass matrix arising from $\mathcal{L}_\text{gf}$.  We prove in appendix \ref{app:VGM} that since they are both built out of the same object, $(gT^a\phi_i)$, the mass matrices $m_A^2(\phi)_{ij}$ and $m_A^2(\phi)^{ab}$ share the same non-zero eigenvalues.

To obtain the one-loop effective potential, we perform the Legendre transform, and the functional integrals are carried out in the gaussian approximation; terms cubic and higher order in quantum fields are dropped from (\ref{gfLag}).  Formally, the result is a sum of functional logarithmic determinant ratios (from now, we drop the `cl' label and bars off the classical background fields):
\begin{multline}\label{effPotDets}
V_\text{eff}(\phi)=V_\text{tree}+\frac{i}{\text{vol}}\bigg[-\frac{1}{2}\ln\left(\frac{\det\mathcal{O}_\text{sc}(\phi)}{\det\mathcal{O}_\text{sc}(0)}\right)\\
-\frac{1}{2}\ln\left(\frac{\det\mathcal{O}_\text{gau}(\phi)}{\det\mathcal{O}_\text{gau}(0)}\right)+\ln\left(\frac{\det\mathcal{O}_\text{FP}(\phi)}{\det\mathcal{O}_\text{FP}(0)}\right)\bigg]\,,
\end{multline}
where
\begin{align}
\nonumber\mathcal{O}_\text{sc}(\phi)&=-\partial^2-M_{ij}^2(\phi)-\xi m_A^a(\phi)_{ij}\\
\nonumber\mathcal{O}_\text{gau}(\phi)&=(\partial^2g^{\mu\nu}-(1-\frac{1}{\xi})\partial^\mu\partial^\nu)\delta^{ab}+m_A^2(\phi)^{ab}g^{\mu\nu}\\
\mathcal{O}_\text{FP}(\phi)&=-\partial^2\delta^{ab}-\xi m_A^2(\phi)^{ab}
\end{align}
are the fluctuation differential operators appearing in (\ref{gfLag}).  The functional determinants may be computed by going to Fourier space, and diagonalizing the mass matrices \cite{Delaunay:2007wb}.  The result is
\begin{multline}\label{effPotTrLog}
V_\text{eff}(\phi)=V_\text{tree}(\phi)\\+\frac{-i}{2}\mu^{2\epsilon}\int\frac{d^dp}{(2\pi)^d}\bigg[\Tr\ln\big(p^2-M_{ij}^2(\phi)-\xi m_A^2(\phi)_{ij}\big)\\
+(d-1)\Tr\ln\big(p^2-m_A^2(\phi)^{ab}\big)+\Tr\ln\big(p^2-\xi m_A^2(\phi)^{ab}\big)\\
-2\, \Tr\ln\big(p^2-\xi m_A^2(\phi)^{ab}\big)-\text{``free''}
\bigg]\,,
\end{multline}
where the UV-divergent momentum integrals are dimensionally regulated ($d=4-2\epsilon$), and ``free'' refers to the field-independent subtractions arising from determinant ratios in (\ref{effPotDets}).  In our notation, the trace-log of a matrix means to sum over the logs of eigenvalues of the matrix.  We note that each term in (\ref{effPotTrLog}) represents fluctuations of specific degrees of freedom in the theory: the first term represents scalar and Goldstone fluctuations; the second arises from the transverse and longitudinal fluctuations of gauge fields, while the third term corresponds to the scalar (time-like) gauge field fluctuations;  and the fourth term represents fluctuations of the Faddeev-Popov ghosts fields.

After adding together the last two logs with identical arguments, we perform the momentum integrals using the standard formula (already expanded in powers of $\epsilon$),
\begin{multline}
\frac{-i}{2}\mu^2\int\frac{d^dp}{(2\pi)^d}\ln(p^2-m^2)=\\
\frac{-1}{4(4\pi)^2}(m^2)^2\left(\frac{1}{\epsilon}-\gamma_E-\ln 4\pi-\ln\left(\frac{m^2}{\mu^2}\right)+\frac{3}{2}\right)
\end{multline}
The $1/\epsilon$ poles are cancelled against appropriate counter-terms by choosing a renormalization scheme.  We choose the convenient $\overline{\text{MS}}$-scheme in which the $1/\epsilon-\gamma_E+\ln 4\pi$ are removed.  The result is
\begin{multline}\label{CWeffPot}
V_\text{eff}(\phi)=V_\text{tree}(\phi)\\+\sum_{\text{scalars},i}\frac{1}{4(4\pi)^2}\left[m_i^2(\phi;\xi)\right]^2\left[\ln\left(\frac{m_i^2(\phi;\xi)}{\mu^2}\right)-\frac{3}{2}\right]\\
+\sum_{\text{gauge},a}\frac{3}{4(4\pi)^2}\left[m_a^2(\phi)\right]^2\left[\ln\left(\frac{m_a^2(\phi)}{\mu^2}\right)-\frac{5}{6}\right]\\
-\sum_{\text{gauge},a}\frac{1}{4(4\pi)^2}\left[\xi m_a^2(\phi)\right]^2\left[\ln\left(\frac{\xi m_a^2(\phi)}{\mu^2}\right)-\frac{3}{2}\right]\,,
\end{multline}
where $m_i^2(\phi;\xi)$ are eigenvalues of the scalar mass matrix $M_{ij}^2(\phi)+\xi m_A^2(\phi)_{ij}$, and $m_a^2(\phi)$ are eigenvalues of the gauge-boson mass matrix $m_A^2(\phi)^{ab}$.  The sums run over these eigenvalues.

\subsection{Origin of gauge dependence}\label{sec:elucidate}
Before generalizing to finite temperature, we pause to comment on the gauge-dependence of the zero-temperature effective potential.  The essential structure is shown in (\ref{effPotTrLog}).  There are three sources of the gauge parameter: from the scalar mass matrix $\xi m_A^2(\phi)_{ij}$ in the first log, from the time-like gauge boson mass matrix $\xi m_A(\phi)^{ab}$ in the third log, and from the Faddeev-Popov mass matrix $\xi m_A(\phi)^{ab}$ in the fourth log.  As $m_A^2(\phi)_{ij}$ and $m_A^2(\phi)^{ab}$ have identical non-vanishing eigenvalues (appendix \ref{app:VGM}), one could in principle arrange a cancellation among these logs by writing
\begin{multline}\label{TrLogSplit}
\Tr\ln\big(p^2-M_{ij}^2(\phi)-\xi m_A^2(\phi)_{ij}\big)\longrightarrow\\
\Tr\ln\big(p^2-M_{ij}^2(\phi)\big)+\Tr\ln\big(p^2-\xi m_A^2(\phi)_{ij}\big)
\end{multline}
provided $m_A^2(\phi)_{ij}$ and $M_{ij}^2(\phi)$ are simultaneously diagonalizable and that their eigenvalues live in distinct subspaces.  However, for general $\phi$, the mass matrices do not decouple, and thus, the effective potential remains gauge-dependent.  We show in Appendix \ref{app:VGM} that in the specific case when $\phi$ extremizes the tree level effective potential, the mass matrices decouple, leading to the necessary cancellation. 

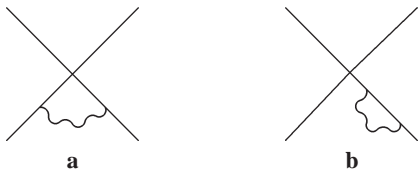
\begin{figure}[t]
\parbox{3cm}{
\begin{fmffile}{ScalarQED1}
\begin{fmfgraph*}(50,50)
  \fmfset{arrow_len}{3mm}
  \fmfpen{0.6pt}
  \fmfstraight
  \fmfleft{i1,i2}
  \fmfright{o1,o2}
  \fmf{plain}{i1,va,v,vb,o1}
  \fmf{plain,tension=0.5}{i2,v,o2}
  \fmffreeze
  \fmf{photon,right=0.5}{va,vb}
\end{fmfgraph*}
\end{fmffile}\\ \textbf{a}}
\qquad
\parbox{3cm}{
\begin{fmffile}{ScalarQED2}
\begin{fmfgraph*}(50,50)
  \fmfset{arrow_len}{3mm}
  \fmfpen{0.6pt}
  \fmfstraight
  \fmfleft{i1,i2}
  \fmfright{o1,o2}
  \fmf{plain}{i1,v}
  \fmf{plain}{v,va}
  \fmf{plain,tension=0.5}{va,vb}
  \fmf{plain}{vb,o1}
  \fmf{plain,tension=0.31}{i2,v}
  \fmf{plain}{v,o2}
  \fmffreeze
  \fmf{photon,right=1}{va,vb}
\end{fmfgraph*}
\end{fmffile}\\\textbf{b}}
\caption{\textbf{a.}\enspace Gauge-dependent scalar-QED type graph describing $\phi\phi$-scattering that contributes to $\Gamma[\phi,A]$.  \textbf{b.}\enspace External-leg corrections necessary to remove gauge-dependence from $\phi\phi$-scattering amplitude, but absent from $\Gamma[\phi,A]$.}
\label{fig:scalarQED}
\end{figure}

The dependence of $V_\text{eff}(\phi)$ on $\xi$ may be traced down to the dependence of the effective action, $\Gamma[\phi,A_\mu]$.   We provide two equivalent ways of understanding the origin of its dependence on $\xi$.  First \cite{Jackiw:1974cv}, the effective action, defined as the sum of 1-PI graphs, includes a scalar QED-type graph shown in Fig. \ref{fig:scalarQED}a, which is gauge dependent.  The graph describes a physical scalar-scalar scattering process that cannot be gauge-dependent.  The sort of graphs needed to cancel gauge dependence from the scattering amplitude is shown in Fig. \ref{fig:scalarQED}b.  Since external leg corrections are not 1-PI graphs, they do not enter the effective action, leading to gauge dependence.

Second, the effective action is equivalently defined as the Legendre transformation of the generating functional of connected Green's functions $W[j]$ with respect to the source $j(x)$, where
\begin{equation}
\label{blah}
W[j]=-i\ln \int \mathcal{D}\phi\,\mathcal{D} A\, \mathcal{D}\eta\, \mathcal{D}\eta^\dag e^{i\int d^dx\big(\mathcal{L}(\xi)+j_i\phi_i\big)}\,.
\end{equation}
Unlike the gauge-fixed Lagrangian, $\mathcal{L}(\xi)$, the source term, $\phi_i j_i$, appearing in the generating functional is not BRST invariant since the source transforms as a gauge singlet.  As a result, the effective action derived from the generating functional in this way will depend on the gauge-fixing procedure, and hence, the gauge parameter.

\subsection{Finite temperature effective potential}\label{sec:finiteT}
We now apply these arguments to the finite temperature effective potential.
The minimizing field $\phi_\text{min}(T)$ and the critical temperature $T_C$ are obtained from the finite temperature effective potential.  The latter is derived from the partition function $Z=\Tr[e^{-H/T}]$ that is expressed as a Euclidean path integral over field configurations satisfying periodic boundary conditions in Euclidean time.

The generalization to finite temperature is achieved by repeating the steps leading to (\ref{effPotTrLog}), but with the energy integral replaced by a sum over Matusbara modes: $\int \frac{dp^0}{2\pi}\longrightarrow T\sum_n$.  The potential is computed in the background field method with the appropriate gauge fixing as before.  We mention that to retain BRST symmetry at finite temperature, the ghost fields must satisfy periodic boundary conditions, despite being Grassmann-valued \cite{Bernard:1974bq}.  After making these modifications, the generalization of (\ref{effPotTrLog}) is
\begin{widetext}
\begin{multline}\label{TeffPotTrLog}
V_\text{eff}(\phi,T)=V_\text{tree}(\phi)+\frac{T}{2}\sum_{n=-\infty}^{+\infty}\mu^{2\epsilon}\int\frac{d^{d-1}\mathbf{p}}{(2\pi)^{d-1}}\bigg[\Tr\ln\left(\omega_n^2+\mathbf{p}^2+M_{ij}^2(\phi)+\xi m_A^2(\phi)_{ij}\right)
\\+(d-1)\Tr\ln\left(\omega_n^2+\mathbf{p}^2+m_A^2(\phi)^{ab}\right)+\Tr\ln\left(\omega_n^2+\mathbf{p}^2+\xi m_A^2(\phi)^{ab})\right)
-2\, \Tr\ln\left(\omega_n^2+\mathbf{p}^2+\xi m_A^2(\phi)^{ab}\right)-\text{``free''}
\bigg]\,,
\end{multline}
where $\omega_n=(2\pi n T)$ are the bosonic Matsubara frequencies and where \lq\lq free" is the free-field subtraction.  After performing the frequency sum, the one-loop contribution separates into a zero-temperature part and a temperature-dependent part,
\begin{equation}\label{finiteTeffPot}
V_\text{eff}(\phi,T)=V_\text{tree}(\phi)+V_\text{CW}(\phi)+\frac{T^4}{2\pi^2}\Big[\sum_{\text{scalar},i}\!\!J_B\left(m_i^2(\phi;\xi)/T^2\right)+3\sum_{\text{gauge},a}\!\!J_B\left(m_a^2(\phi)/T^2\right)-\sum_{\text{gauge},a}\!\!J_B\left(\xi m_a^2(\phi)/T^2\right)\Big]\,,
\end{equation}
where $V_\text{tree}(\phi)+V_\text{CW}(\phi)$ is the Coleman-Weinberg effective potential in Eq.~(\ref{CWeffPot}), $J_B(z^2)=\int_0^\infty dx\,x^2\ln(1-e^{-\sqrt{x^2+z^2}})$ is the bosonic thermal function, and where $m_i^2(\phi;\xi)$ and $m_a^2(\phi)$ are, respectively, the eigenvalues of the scalar and gauge boson mass matrices as defined below Eq.~(\ref{CWeffPot}).  The thermal functions are typically numerically integrated or approximated using a Bessel function representation.
\end{widetext}

Clearly, the finite-temperature effective potential (\ref{finiteTeffPot}) is gauge dependent.  The dependence appears for precisely the same reasons described in the previous section.  The scalar trace-log represented by the first term under the integral in Eq.~(\ref{TeffPotTrLog}) cannot be split in a manner similar to Eq.~(\ref{TrLogSplit}) because the scalar mass matrices $M_{ij}^2(\phi)$ and $m_A^2(\phi)_{ij}$ are generally not simultaneously diagonalizable with eigenvalues living in distinct subspaces.

\subsection{Implications for the standard analysis}
We now show that the gauge-dependence of the effective potential leads to an unphysical treatment of the BNPC in the standard analysis.  For the sake of clarity we temporarily restrict our discussion to the SM.

The critical temperature $T_C$ marking a phase transition between two phases $\phi^{(1)}_i$ and $\phi^{(2)}_i$, derived from (\ref{finiteTeffPot}), is defined by
\begin{gather}
\label{defDegen} V_\text{eff}(\phi^{(1)}_i,T_C;\xi)-V_\text{eff}(\phi_i^{(2)},T_C;\xi)=0\\
\label{defMin} \frac{\partial V_\text{eff}}{\partial\phi_i}\Big|_{\phi^{(1)}_i,T_C}=\frac{\partial V_\text{eff}}{\partial\phi_i}\Big|_{\phi^{(2)}_i,T_C}=0\,.
\end{gather}
 In the SM, $\phi^{(1)}=0$ is the symmetric phase, while $\phi^{(2)}\equiv \phi_\text{min}(T_C)$ characterizes the broken phase. In the standard analysis, (\ref{defDegen}) and (\ref{defMin}) are simultaneously solved to obtain both $T_C$ and $\phi_\text{min}(T_C)$.   Their ratio is compared to the bound (\ref{eq:exp}) required for preservation of the baryon asymmetry.

It is straightforward to see that na\"{i}vely inverting these equations at one-loop order leads to a gauge-dependent estimate of the sphaleron rate.  To that end, we consider the high-$T$ approximation to the full effective potential. The thermal bosonic function $J_B(z^2)$ admits a high-temperature expansion \cite{Dolan:1973qd}
\begin{equation}\label{eq:highTbosonic}
J_B(z^2)=-\frac{\pi^4}{45}+\frac{\pi^2}{12}z^2-\frac{\pi}{6}(z^2)^{3/2}-\frac{1}{32}z^4\ln z^2+\ldots,
\end{equation}
with which the effective potential (\ref{finiteTeffPot}) may be cast as a polynomial in $\phi$:
\begin{equation}\label{highTexp}
V_\text{eff}(\phi,T)=D(T^2-T_0^2)\phi^2-ET\phi^3 +\frac{\bar\lambda}{4} \phi^4+\ldots.
\end{equation}
The coefficients $D$, $T_0^2$, $E$ and $\bar\lambda$ depend on the parameters of the underlying model.  In the SM, the coefficients are \cite{Anderson:1991zb}
\begin{equation}
\label{eq:coeffdef}
\begin{aligned}
D&=\frac{1}{32}(g_1^2+3g_2^2+4y_t^2+8\lambda)\,,\\
T_0^2&=\mu^2/2D\,,\\
E&=\frac{3-\xi^{3/2}}{96\pi}\big(2 g_2^3+(g_1^2+g_2^2)^{3/2}\big)\,,\\
\text{and}\enspace\bar\lambda&=\lambda+(\text{$\xi$-dep. log})\,,
\end{aligned}
\end{equation}
where $y_t$ is the top yukawa coupling; $g_1$ and $g_2$ are the U(1) and SU(2$)_L$ gauge coupling constants; and the scalar quartic self coupling $\bar \lambda$ picks up a logarithmic $\xi$-dependence. 

We observe that the coefficient of the quadratic term is gauge-independent, as one expects based on the gauge-independence of thermal masses (see {\em e.g.}, Ref.~\cite{Braaten:1989kk}).  In Appendix \ref{app:highTexp}, we explicitly demonstrate this property for the general model.  As we will discuss below, we take advantage of this property to define the high-temperature effective theory used to obtain a gauge-independent sphaleron scale.

Unfortunately, the coefficient $E$ is not only gauge-dependent but strongly so.  For example, by choosing $\xi=3^{2/3}$ the $E$-coefficient can be made to vanish, and the barrier necessary for a first order phase transition is permanently absent.  One might hope that the ratio $\phi_\text{min}(T_C)/T_C$ removes this gauge-dependence. However it is straightforward to show that this hope is not realized. A simple calculation gives (for a pedagogical review, see Ref.~\cite{Quiros:1999jp})
\begin{equation}
\frac{\phi_\text{min}(T_C)}{T_C}=\frac{2E}{\bar\lambda}\,,
\end{equation}
which remains gauge dependent.  Hence the standard analysis leads to an unphysical treatment of the BNPC.

In various Standard Model extensions where either new scalar loops or tree-level operators generate large contributions to the cubic term, it is conceivable that the impact of the gauge dependence is numerically minimized so long as one works in a region near $\xi=0$. Nevertheless, such a situation would be wholly unsatisfying as one would not know whether a perturbative computation near $\xi=0$ reflects reality or introduces a substantial gauge-dependent artifact. Instead, one should endeavor to maintain strict gauge-independence throughout the analysis in order to avoid any ambiguity.  Before moving on to our proposal for a gauge-independent analysis, we briefly review some theoretical properties of the effective potential that will prove useful later. 

\subsection{Nielsen identities}\label{sec:Nielsen}
The dependence of the effective action on the gauge-fixing parameter is described by the Nielsen identity \cite{Nielsen:1975fs} \cite{Fukuda:1975di}, and its precise form depends on the gauge fixing condition imposed on the quantum fields.  The identity  follows directly from the BRST (non)invariance of the sourced generating functional.  For the general class of linear gauges, the identity reads
\begin{multline}\label{nielsenAction}
\frac{\partial \Gamma}{\partial \xi}=\int d^dx\, d^dy\Big[C_i(\phi,A;\,x,y)\frac{\delta\Gamma}{\delta\phi_i(x)}\\
+E^a_\mu(\phi,A;\,x,y)\frac{\delta\Gamma}{\delta A_\mu^a(x)}\Big]\,,
\end{multline}
where, $C$ and $E$ stand for field-dependent vacuum correlators.  The interpretation of the identity is clear: the value of the effective action evaluated at its stationary points, i.e. at points where the classical fields satisfy $\delta\Gamma/\delta\phi_i(x)=\delta\Gamma/\delta A_\mu^a(x)=0$, is gauge-independent.  This is as it should be since physical quantities are derived from the stationary points of the effective action \cite{Thompson:1985hp}.

\begin{figure}
\includegraphics[scale=1,width=6cm]{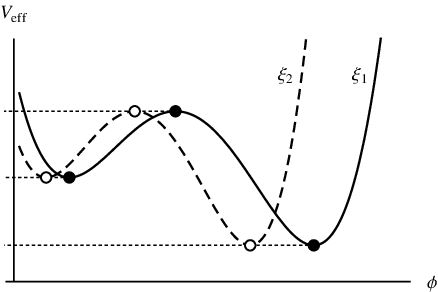}
\caption{A schematic illustration of the behavior of the exact effective potential as the gauge parameter $\xi$ is varied according to Nielsen's identity.  The values of the potential at its extrema stay unchanged but the fields extremizing the potential are gauge-dependent.}\label{fig:nielSketch}
\end{figure}

To compute the effective potential, the vector potentials are taken to vanish $A_\mu(x)=0$, and the scalar fields assumed to be homogenous $\phi(x)\equiv\phi$.  In this case, the effective action reduces to the effective potential, $\Gamma[\phi]/\mathrm{vol}= -V_\text{eff}(\phi)$, and the Nielsen identity simplifies to
\begin{equation}\label{nielsenPotential}
\frac{\partial V_\text{eff}}{\partial\xi}=-C_i(\phi,\xi)\frac{\partial V_\text{eff}}{\partial\phi_i}\,,
\end{equation}
which carries the same interpretation: the effective potential is gauge-independent where it is stationary.  Note, however, that the point in field space minimizing (or extremizing) the effective potential $\phi_\text{min}(\xi)$ is not gauge independent; as the gauge parameter is varied, the effective potential compresses or expands along the $\phi$-axis (see Fig. \ref{fig:nielSketch}), while maintaining gauge independence of the energy density at its extremal points.

The derivation of (\ref{nielsenAction}) and (\ref{nielsenPotential}) does not rely on performing any non-trivial space-time or energy-momentum integrations.  Thus, Nielsen's identities are valid at finite temperature upon the standard replacement  $\int \frac{dp^0}{2\pi}\longrightarrow T\sum_n$.  An immediate consequence, proven in Appendix \ref{app:critTemp}, is the gauge-independence of the critical temperature.  However, as mentioned above, the precise value of the minimizing field $\phi(T_C)$ remains gauge dependent (see Appendix \ref{app:scCond} for details).  Thus, the ratio $\phi(T_C)/T_C$ conventionally used to establish the BNPC is inherently gauge dependent.

\subsection{Artificial violations of Nielsen's identities}
Although gauge independence of physical quantities is an exact statement following from Nielsen's identities, gauge dependence may be introduced to physical quantities as an artifact of an approximation scheme.  In particular, na\"ively truncating the perturbative effective potential at a finite order in perturbation theory leads to apparent violations of Nielsen's identity.  In fact, we identify this na\"ive truncation as the principal source of gauge dependence in the standard determination of $T_C$.

Such effects may be deduced directly from Nielsen's identity.  By expressing $V_\text{eff}(\phi)$ and $C(\phi,\xi)$ in (\ref{nielsenPotential}) as a series in $\hbar$,
\begin{align}
V_\text{eff}(\phi)&=V_0(\phi)+\hbar\,V_1(\phi)+\hbar^2\,V_2(\phi)+\ldots\\
C(\phi,\xi)&=c_0+\hbar\,c_1(\phi)+\hbar^2\,c_2(\phi)+\ldots\,,
\end{align}
and retaining terms through $\mathcal{O}(\hbar)$ in both sides of (\ref{nielsenPotential}), we find
\begin{equation}
\frac{\partial V_0}{\partial\xi}+\hbar\frac{\partial V_1}{\partial\xi}=-c_0\frac{\partial V_0}{\partial\phi}-\hbar\Big(c_0\frac{\partial V_1}{\partial\phi}+c_1\frac{\partial V_0}{\partial \phi}\Big)\,.
\end{equation}
Since the tree-level potential $V_0$ is strictly gauge independent, setting $\mathcal{O}(\hbar^0)$ terms equal implies $c_0=0$.  Upon setting $\mathcal{O}(\hbar^1)$ terms equal we have
\begin{equation}
\frac{\partial V_1}{\partial \xi}=-c_1\frac{\partial V_0}{\partial\phi}\,;
\end{equation}
that is, the one-loop potential is gauge-independent only where the tree-level potential is extremized, and not where the one-loop potential is extremized.  Therefore, the critical temperature based on the one-loop extremum is gauge-dependent.  We note that gauge dependence appears formally at a higher order than the order of approximation.  Yet, numerically, there is no limit to its sensitivity, and one should necessarily strive to obtain a critical temperature that strictly maintains gauge-independence at each order in perturbation theory.

\section{Gauge-independent determination of $T_C$}\label{GIanalysisSection}
In the following sections, we detail one method for performing a gauge-independent perturbative analysis of the EWPT. As discussed above, exact expressions for the critical temperature and sphaleron rate are not at our disposal; each has to be computed in an approximation scheme.  We focus on the critical temperature in this section, and discuss the sphaleron scale in the next section.  We stress that the computation of the critical temperature and the temperature-dependent sphaleron energy are two independent calculations, and that each should separately be gauge-independent.  Since they are derived from different quantities -- one from the effective potential, the other from the effective action -- the analysis should be treated as such, by computing both quantities separately.

\subsection{The $\hbar$ expansion}\label{sec:GIcritT}
The key to extracting a gauge-independent critical temperature is to invert the defining equations (\ref{defDegen}) and (\ref{defMin}) in a manner consistent with the $\hbar$-expansion.  To that end, we start by computing the effective potential to the desired order in perturbation theory as outlined in Sec. \ref{sec:GaugeProblem}:
\begin{equation}\label{Vexpanded}
V_\text{eff}(\phi,T)=V_0(\phi)+\hbar\,V_1(\phi,T)+\hbar^2\,V_2(\phi,T)+\ldots
\end{equation}
We wish to minimize this function,
\begin{equation}\label{Vmincond}
\frac{\partial V_\text{eff}}{\partial \phi}\Big|_{\phi_\text{min}}=0\,.
\end{equation}
We then write $\phi_\text{min}$ as a series in $\hbar$
\begin{equation}
\phi_\text{min}=\phi_0+\hbar\,\phi_1(T,\xi)+\hbar^2\,\phi_2(T,\xi)+\ldots\,,
\end{equation}
where $\phi_0$ represents any one of the minima of the tree-level effective potential.
Upon substitution into (\ref{Vmincond}), we find
\begin{align}\nonumber
\frac{\partial V_0}{\partial \phi}\Big|_{\phi_0+\hbar\phi_1+\ldots}+\hbar\frac{\partial V_1}{\partial \phi}\Big|_{\phi_0+\hbar\phi_1+\ldots}+\ldots&=0\\
\frac{\partial V_0}{\partial\phi}\Big|_{\phi_0}+\hbar\Big(\frac{\partial V_1}{\partial \phi}\Big|_{\phi_0}+\phi_1\frac{\partial^2 V_0}{\partial\phi^2}\Big|_{\phi_0}\Big)+\mathcal{O}(\hbar^2)&=0\,,
\end{align}
where in the second line, we have expanded and organized in powers of $\hbar$.  The requirement that each coefficient must vanish individually determines the position of the minimum $\phi_\text{min}$ at each order in perturbation theory:
\begin{equation}\label{eq:condensate}
\begin{aligned}
\mathcal{O}(\hbar^0):&&0&=\frac{\partial V_0}{\partial \phi}\Big|_{\phi_0}\\
\mathcal{O}(\hbar^1):&&\phi_1(T,\xi)&=-\Big(\frac{\partial^2 V_0}{\partial\phi^2}\Big)^{-1}_{\phi_0}\frac{\partial V_1(T,\xi)}{\partial\phi}\Big|_{\phi_0}
\end{aligned}
\end{equation}
Here, $\phi_1(T,\xi)$ represents the genuine, albeit gauge-dependent, one-loop correction to the tree-level VEV $\phi_0$ at zero temperature, or to the background field at finite temperature.  After substituting these back into (\ref{Vexpanded}) and expanding again in $\hbar$, we find
\begin{multline}\label{Vcurve}
V_\text{eff}(\phi_\text{min}(T),\,T)=V_0(\phi_0)+\hbar V_1(\phi_0,T)\\
+\hbar^2\big[V_2(\phi_0,T,\xi)-\textstyle\frac{1}{2}\phi_1^2(T,\xi) \frac{\partial^2 V_0}{\partial\phi^2}|_{\phi_0}\big]+\mathcal{O}(\hbar^3)\,.
\end{multline}
This formula for $V_\text{eff}(\phi_\text{min}(T),T)$ appears in Ref.~\cite{Laine:1994zq}, applied to the high-temperature effective theory.

At each order in $\hbar$, the RHS of Eq.~(\ref{Vcurve}) is gauge independent in accordance with Nielsen's identity.  An explicit verification of this for the $\mathcal{O}(\hbar)$ term is as follows.  The one-loop expression is shown in (\ref{TeffPotTrLog}) and must be evaluated at the tree level minimum $\phi_0$.  Thus, the mass matrices $M_{ij}^2(\phi_0)$ and $m_A^2(\phi_0)_{ij}$ in (\ref{TeffPotTrLog}) are evaluated at $\phi_0$.  At that point, these mass matrices are simultaneously diagonalizable and eigenvalues live in distinct subspaces (Appendix \ref{app:VGM}). As discussed in section \ref{sec:elucidate}, this situation allows us to split the scalar log into two terms as in (\ref{TrLogSplit}) and finally achieve the needed cancellation among the gauge-dependent logarithms.

\begin{figure}
\includegraphics[scale=1,width=7cm]{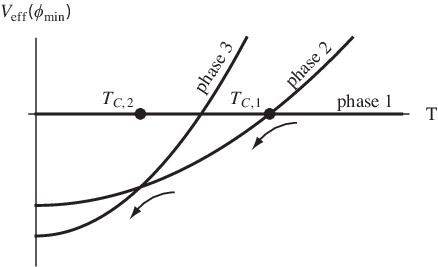}
\caption{Value of $V_\text{eff}(\phi_\text{min})$ for three different phases as a function of $T$ as determined by (\ref{Vcurve}) in a hypothetical theory.  Evolution of the universe proceeds from right to left, following the direction of the arrows.  Intersections of the two lowest curves define the critical temperatures (there are two in this example).}\label{fig:Vcurve}
\end{figure}

Equation (\ref{Vcurve}) provides a gauge-independent closed-form expression for the value of the effective potential at its local minimum (or maximum) as a function of temperature.  Since the function may be computed from any extremum $\phi_0$ derived from the tree-level potential $V_0$, equation (\ref{Vcurve}) defines a family of curves which follow the temperature evolution of the extremal points.  At each intersection point, the degeneracy condition (\ref{defDegen}) is satisfied, and that point potentially corresponds to a critical temperature.  Since phase transitions typically take the system from one global minimum of $V_\text{eff}$ to another, critical temperatures occur when the two lowest curves intersect.  Fig. \ref{fig:Vcurve} depicts a scenario for a hypothetical theory with three phases.  Evolution of the universe proceeds from right to left.  In this example, the phase transition occurs in two steps.  First, a transition occurs from phase 1 to phase 2 at temperature $T_{C,\,1}$.  Then a second transition takes place from phase 2 to phase 3 at temperature $T_{C,\,2}$.

Note that, in stark contrast with the current analysis which focuses on the complicated multivariable function $V_\text{eff}(\phi,T)$ to follow the free energy of each phase, the foregoing method deals with a family of functions dependent on the single variable $T$.  Because numerical algorithms that find intersection points of two curves are significantly more efficient than those that find global minima of multivariable functions, our method to find critical temperatures is substantially faster than the standard methods.  This is an added benefit of the gauge-independent approach discussed above, as it would speed up the analysis of multi-step phase transitions in complicated extensions of the SM that are subject to extensive parameter scans.

\subsection{Qualitative analysis of the gauge-independent critical temperature}
One may get a sense of the behavior of the critical temperature at $\mathcal{O}(\hbar)$ by analyzing (\ref{Vcurve}) in the high-$T$ limit.  We shall restrict ourselves to a single degree of freedom and study a phase transition from the symmetric phase to the broken phase.  We normalize the potential such that it vanishes at the origin at all temperatures, so that the degeneracy condition (\ref{defDegen}) to $\mathcal{O}(\hbar)$ reads
\begin{equation}\label{eq:degenV1}
V_\text{eff}(\phi,T_C)=V_0(\phi_0)+\hbar V_1(\phi_0,T)=0\,.
\end{equation}
After applying the high-temperature expansion of the thermal bosonic function (\ref{eq:highTbosonic}) to the LHS of (\ref{eq:degenV1}) we have
\begin{equation}
D(T^2-T_0^2)\phi_0^2-ET\phi_0^3+\frac{\bar\lambda}{4}\phi_0^4=0\,.
\end{equation}
Note that gauge-independence of this expression is ensured only when the fields are evaluated at their tree-level minima $\phi_0=T_0\sqrt{2D/\bar\lambda}$ as dictated by the $\hbar$-expansion.  Solving this equation for $T$ yields two roots; only one gives a positive critical temperature
\begin{equation}
T_C=T_0\bigg(\frac{E T_0}{\bar\lambda \phi_0}+\sqrt{\bigg(\frac{E T_0}{\bar\lambda \phi_0}\bigg)^2+\frac{1}{2}}\bigg)\,,
\end{equation}
where $D$ was eliminated in favor of $\phi_0$.  Assuming no tree-level cubic terms, we may take $E$ small, since it is loop-suppressed.  We the find
\begin{equation}
T_C=\frac{T_0}{\sqrt{2}}\approx 0.71\, T_0\,,
\end{equation}
that the critical temperature is approximately 30\% lower than $T_0$ (the critical temperature in the second-order limit).  Our  one-loop numerical study of $T_C$ presented in Section \ref{sec:SM} confirms these expectations.  Writing $T_0=\phi_0\sqrt{\bar\lambda/2D}$, we find the scaling behavior of the critical temperature
\begin{equation}\label{eq:giTrend}
T_C=\frac{\phi_0}{2}\sqrt{\frac{\bar\lambda}{D}}\,.
\end{equation}
As the quartic coupling $\bar\lambda$ is increased, the critical temperature rises, too.

\subsection{Higher-order contributions}
When analyzing the behavior of the effective potential, it is instructive to investigate the impact of higher-order contributions and to explore the limits of validity of perturbation theory.  In this section we study these higher-order contributions to assess their importance to (\ref{Vcurve}).

It is well-known that at high temperatures the perturbative expansion breaks down due to uncontrolled $g^2T^2/m^2$ terms appearing at higher orders\cite{Dolan:1973qd}\cite{Linde:1978px}\cite{Gross:1980br}.  These \lq\lq plasma damping" contributions are induced by fluctuations of  $n>0$ Matsubara modes at high temperature.  This malady is cured by re-summing these terms to all orders to obtain a ``ring-improved'' effective potential $V_\text{eff}^\text{ring}(\phi,T)$ that includes the damping effects. In this subsection, we will discuss two possible gauge-independent methods of including such effects.  We will then conclude by briefly discussing the $\mathcal{O}(\hbar^2)$ term of (\ref{Vcurve}).

\subsubsection{High-$T$ effective theory}
One may include damping effects by going to the high-$T$ effective theory by integrating out the heavy Matsubara modes via dimensional reduction.  Nielsen's identities (\ref{nielsenAction}) and (\ref{nielsenPotential}) apply to any gauge theory, including effective theories.  Provided no explicit gauge-dependence is introduced in the Wilson coefficients of the effective theory, stationary points of the effective potential derived from it will remain gauge-independent.  For the same reasons that the effective action is gauge-dependent, so is the effective potential obtained with the dimensional reduction procedure.  Indeed, the Wilson coefficients of all terms, apart from the $\mathcal{O}(T^2)$ terms are gauge-dependent.  However, by working in the limit $T\gg m_W^{T=0}$, one can justify retaining only these $\mathcal{O}(T^2)$ terms of the effective theory.  These terms are gauge independent because thermal masses are gauge independent.  In principle, then, it is possible within this limit to derive a gauge-independent critical temperature from the high-$T$ effective theory with the $g^2T^2/m^2$ terms implicitly re-summed.

This procedure has been applied to the SU(2) Higgs model in Ref.~\cite{Laine:1994zq} , where it was argued that the $\hbar$ expansion for $T_C$ breaks down at two-loop order.  We revisit this analysis briefly below and also conclude that this approach is not useful for obtaining information about the phase transition. 

To proceed, we start with the effective lagrangian, where the heavy Matsubara modes are integrated out\cite{Laine:1994zq,Farakos:1994kx,Jakovac:1994mq}: 
\begin{multline}
\mathcal{L}_3=\frac{1}{2}(\vec{D}\Phi_i)^2+V(\Phi)+\frac{1}{2}\Sigma_{ij}(T)\,\Phi_i\Phi_j\\
+\frac{1}{4}\tensor{F}^a.\tensor{F}^a+\frac{1}{2} (\vec{D} A_0^a)^2+\frac{1}{2}\Pi^{ab}(T)\,A_0^aA_0^b\\
+\Lambda^{abcd} A_0^a A_0^b A_0^c A_0^d+H^{ab}_{ij} A_0^a A_0^b \Phi_i \Phi_j
+j_i\Phi_i\,,
\end{multline}
where $\tensor{F}$ is the 3D field strength tensor, $\Sigma(T)$ and $\Pi(T)$ are scalar and gauge-boson thermal mass matrices, respectively;  $H_{ij}^{ab}$ gives the tree-level quartic coupling of the scalar and time-like gauge fields; and $\Lambda^{abcd}$ is a $T$-independent loop induced quartic interaction of the time-like gauge fields. Note that we have not yet performed the integration over the $n=0$ Matsubara modes.  In this theory, the tree-level scalar potential is
\begin{equation}\label{eq:VhighTeff0}
V_0^\text{high-$T$}(\Phi,T)=V(\Phi)+\frac{1}{2}\Sigma_{ij}(T)\Phi_i\Phi_j\,.
\end{equation}
In the standard SU(2) model with a single degree of freedom, this potential may be written in a form similar to the high temperature-expansion of the effective potential derived from the full theory (\ref{highTexp}):
\begin{equation}
\label{eq:VhighTeff}
V_0^\text{high-$T$}(\phi,T)=D(T^2-T_0^2)\phi^2+\frac{\bar{\lambda}}{4}\phi^4\,,
\end{equation}
where $DT_0^2=-\mu^2/2$ and where $DT^2$ incorporates the effect of the scalar thermal mass with $D$ being given in (\ref{eq:coeffdef}).
In contrast to the high-$T$ expansion of the full gauge-dependent effective potential (\ref{highTexp}), the RHS of Eq.~(\ref{eq:VhighTeff}) contains 
no $T\phi^3$ term. Its appearance in (\ref{highTexp}) is due to Matsubara zero modes, which we have not yet integrated out to derive (\ref{eq:VhighTeff}).  Thus, at tree level, the phase transition is second order and the critical temperature is $T_0$.

The one-loop effective potential is computed following the background field method, with an $R_\xi$ gauge fixing condition similar to (\ref{eq:gfCondition}).  Upon performing the functional integration over the zero modes, the high-$T$ effective potential acquires the non-analytic terms necessary for a first order phase transition.
\begin{multline}
V_\text{eff}^\text{high-$T$}(\phi,T)=V_0^\text{high-$T$}(\phi,T)\\
-\frac{T}{12\pi}\sum_\text{e-values}\bigg[\big(M_{ij}^2(\phi)+\xi m_A^2(\phi)_{ij}+\Sigma_{ij}(T)\big)^{3/2}\\
+2\big(m_A^2(\phi)^{ab}\big)^{3/2}+\big(m_A^2(\phi)^{ab}+\Pi^{ab}(T)\big)^{3/2}\\
\label{eq:vring0}
-\big(\xi m_A^2(\phi)^{ab}\big)^{3/2}\bigg]\,,
\end{multline}
where the sum runs over the eigenvalues of the field-dependent mass matrices defined in below (\ref{homBGlag}) and (\ref{gfLag}).

The potential is gauge dependent as expected.  Nevertheless, one might expect to obtain a gauge independent critical temperature by applying the $\hbar$-expansion (\ref{Vcurve}) on the high-$T$ effective potential and determining intersection points as outlined section \ref{sec:GIcritT}.  Indeed, the cancellation of gauge-dependence operates in a manner analogous to that in the full 4D theory.  The difference in this case is the tree-level extrema derived from (\ref{eq:VhighTeff0}) are now temperature dependent $\phi_0\equiv\phi_0(T)$.  So, when the scalar mass matrices are evaluated at these points in accordance with (\ref{Vcurve}), the eigenvalues of $M_{ij}^2\big(\phi_0(T)\big)+\Sigma_{ij}(T)$ and $\xi m_A^2\big(\phi_0(T)\big)_{ij}$ decouple, allowing us to write
\begin{multline}\label{eq:splitCube}
\big[M_{ij}^2\big(\phi_0(T)\big)+\xi m_A^2\big(\phi_0(T)\big)_{ij}+\Sigma_{ij}(T)\big]^{3/2}\longrightarrow\\
\big[M_{ij}^2\big(\phi_0(T)\big)+\Sigma_{ij}(T)\big]^{3/2}+\big[\xi m_A^2\big(\phi_0(T)\big)_{ij}\big]^{3/2}\,,
\end{multline}
and arrange a cancellation with the last term of (\ref{eq:vring0}) after applying theorem 2 in Appendix \ref{app:VGM}.

Before discussing the physical implications of the above procedure, we contrast this method with the approach taken in Ref.~\cite{Laine:1994zq}, which, in addition to $V_\text{eff}$ and $\phi_\text{min}$, $T_C$ is itself written in powers of $\hbar$:
\begin{equation}
T_C=T_0+\hbar T_1+\hbar^2 T_2+\ldots\,.
\end{equation}
As before, the leading order critical temperature is $T_0$ and the transition is second order.  Inclusion of the $\mathcal{O}(\hbar)$ terms yields a contribution $T_1$ proportional to the product\footnote{In the notation of \cite{Laine:1994zq} these are denoted $h_3$ and $m_D$, respectively.} of $H^{ab}_{ij}$ and $\Pi^{ab}(T)$.  At $\mathcal{O}(\hbar^2)$, the scalar mass acquires a logarithmic dependence on the renormalization scale $\mu$ as needed to maintain the renormalization group invariance of the potential. Ref.~\cite{Laine:1994zq} finds that this is associated with a two-loop contribution to the potential of the form
\begin{equation}
\label{eq:logdiv}
\ln\left(\frac{D(T^2-T_0^2)}{\mu^2}\right)\, .
\end{equation}
When solving for $T_2$, the argument of the logarithm must be evaluated at $T=T_0$ in order to maintain consistency with the $\hbar$-expansion of $T_C$. In this case, the argument of the logarithm vanishes, signaling the presence of an IR divergence. As Ref.~\cite{Laine:1994zq} concludes, the $\hbar$ expansion clearly fails when implemented in this manner. 

We believe that our method proposed in section \ref{sec:GIcritT} does not suffer from this breakdown.  Note that our method of determining the critical temperature just requires the analysis of the $\hbar$ series expansion of $V_\text{eff}(\phi_\text{min})$ in (\ref{Vcurve}). Since this quantity is already gauge-independent at any value of the temperature, a series expansion of $T_C$ is not needed.  Instead we take the temperature as an external parameter and then numerically solve for $T_C$ that satisfies the degeneracy condition (\ref{defDegen}).  In general, then, the value of $T_C$ appearing in the argument of the logarithm in Eq.~(\ref{eq:logdiv}) will differ from its leading order value $T_0$, and so one should not encounter an IR divergence with this method. One must of course separately treat the specific case when the solution of the degeneracy condition gives $T_C=T_0$ exactly.  As we also note below, we do not encounter any IR divergences at $\mathcal{O}(\hbar^2)$ when determining $T_C$ using the full theory in the high-$T$ regime. Thus, it appears to us that the breakdown of the $\hbar$ expansion as obtained in Ref.~\cite{Laine:1994zq} is a consequence of expanding $T_C$ itself.

We emphasize that these observations do not constitute a criticism of the work of Ref.~\cite{Laine:1994zq}. In that work, the full $\mathcal{O}(\hbar^2)$ contributions to the minima of the effective potential in the dimensionally reduced effective theory were computed explicitly and shown to be gauge-independent and an expression for the value that minimum similar to Eq.~(\ref{Vcurve}) appears. Moreover, a careful discussion of IR divergences for $T$ in the vicinity of $T_0$ was given. We merely differ on the application of the $\hbar$-expansion to $T_C$ itself. 

That being said, we also conclude that this approach using the dimensionally-reduced, high-$T$ effective theory is unlikely to yield useful information about the true critical temperature. The reason is that in the dimensionally reduced theory, at leading order in $\hbar$ in the SM, there exists only one minimum at any given temperature. At $T>T_0$, this minimum lies at $\phi_0=0$. For $T<T_0$, $\phi_0=0$ gives a local maximum while $\phi_0^2=2 D(T_0^2-T^2)/\lambda$. At $T=T_0$ the two coincide, and one encounters the IR divergence when evaluating the higher order terms at $\phi_0=0$. It is unlikely that for $T<T_0$ the higher order corrections convert the tree-level local maximum at the origin into a local minimum, so that one expects the only critical temperature that ever appears is $T_0$. Thus, we abandon this approach for incorporating the plasma damping effects into the determination of $T_C$. 

\subsubsection{A ring-sum prescription}\label{sec:ringsumpres}

\begin{figure}
\includegraphics[scale=.3,width=3cm]{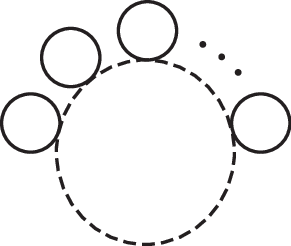}
\caption{A representative ring diagram contributing to $\Delta V_\text{ring}(\phi,T)$.  The Matsubara $n=0$ mode propagates in the lower loop.  The smaller loops are hard thermal loops, each of which contributes a factor of $g^2T^2/m^2$.}
\label{fig:Ring1}
\end{figure}

The second alternative to including the damping effects entails a variant on the conventional approach based on the full 4D theory.  In this approach, the $g^2T^2/m^2$ terms are understood to arise from the high temperature limit of ring (or daisy) graphs shown in Fig. \ref{fig:Ring1}.  The summation of these graphs to all orders results in an additional contribution to the potential (\ref{finiteTeffPot}) of the form
\begin{equation}\label{eq:ringV}
\Delta V_\text{ring}(\phi,T)=V_\text{ring}^{[B]}(\phi,T)-V_\text{ring}^{[A]}(\phi,T)\,,
\end{equation}
where
\begin{multline}
V_\text{ring}^{[A]}(\phi,T)=-\hbar\frac{T}{12\pi} \sum_{\text{e-values}}\Big[\big(M_{ij}^2(\phi)+\xi m_A^2(\phi)_{ij}\big)^{3/2}\\
+3\big(m_A^2(\phi)^{ab}\big)^{3/2}-\big(\xi m_A^2(\phi)^{ab}\big)^{3/2}\Big]\,,
\end{multline}
is the non-analytic part of $V_\text{eff}(\phi,T)$ in (\ref{eq:highTexpGen}) and gets replaced by the thermal damping expression
\begin{multline}
V_\text{ring}^{[B]}(\phi,T)=\\
-\hbar\frac{T}{12\pi}\sum_\text{e-values}\bigg[\big(M_{ij}^2(\phi)+\xi m_A^2(\phi)_{ij}+\Sigma_{ij}(T)\big)^{3/2}\\
+2\big(m_A^2(\phi)^{ab}\big)^{3/2}+\big(m_A^2(\phi)^{ab}+\Pi^{ab}(T)\big)^{3/2}\\
-\big(\xi m_A^2(\phi)^{ab}\big)^{3/2}\bigg]\,,
\end{multline}
which coincides with the 1-loop expression of $V_\text{eff}^\text{high-$T$}(\phi,T)$ in (\ref{eq:vring0}). 
These may be added to bring $\Delta V_\text{ring}(\phi,T)$ into a more recognizable form
\begin{multline}
\label{eq:vring1}
\Delta V_\text{ring}(\phi,T)=\\
-\hbar\frac{T}{12\pi}\sum_{\text{scalars, }i}\Big[\big(m^2(\phi,\xi)+\Sigma(T)\big)_i^{3/2}-m_i^2(\phi;\xi)^{3/2}\Big]\\
-\hbar\frac{T}{12\pi}\sum_{\substack{\text{longit.}\\\text{gauge, }a}}\Big[\big(m^2_A(\phi)+\Pi(T)\big)_a^{3/2}-m_A^2(\phi)_a^{3/2}\Big]\,,
\end{multline}
where $m^2(\phi;\xi)=M^2(\phi)+\xi m_A(\phi)$ is the scalar mass matrix, $m_A(\phi)$ is the gauge boson mass matrix, and $\Sigma(T)$ and $\Pi(T)$ are their gauge-independent thermal self-energy matrices which start at $\mathcal{O}(\hbar)$.  The sums run over the eigenvalues of these matrices.

Note that since $\Sigma(T)$ and $\Pi(T)$ are $\mathcal{O}(\hbar)$, the procedure of re-summing these graphs necessarily departs from the strict $\hbar$-expansion needed to ensure gauge-independence. Thus, it is a non-trivial problem to implement the re-summation into the gauge-independent analysis as one would have to analyze the all-orders expression of (\ref{Vcurve}) and identify the $g^2T^2/m^2$ terms that need to be re-summed.

Instead we provide a gauge-independent prescription for its inclusion based on the following observation.  Earlier, we pointed out that $V_\text{ring}^{[B]}(\phi,T)$ is just the one-loop expression of the effective potential derived from the high-$T$ theory, and $V_\text{ring}^{[A]}(\phi,T)$ is the $\mathcal{O}(T)$ part of the effective potential derived from the full 4D theory.  Thus, if we evaluate $V_\text{ring}^{[B]}$ at $\phi_0(T)$, which minimizes the tree-level high-$T$ potential $V_0^\text{high-$T$}(\phi,T)$ in (\ref{eq:VhighTeff0}), we may apply (\ref{eq:splitCube}) to cancel the gauge-dependence in $V_\text{ring}^{[B]}(\phi_0(T),T)$.  Similarly, if we evaluate $V_\text{ring}^{[A]}$ at $\phi_0$, which minimizes the tree-level potential of the full 4D theory (\ref{VGM}), we may apply
\begin{multline*}
\big(M_{ij}^2(\phi_0)+\xi m_A^2(\phi_0)_{ij}\big)^{3/2}\longrightarrow\\
\big(M_{ij}^2(\phi_0)\big)^{3/2}+\big(\xi m_A^2(\phi_0)_{ij}\big)^{3/2}\,,
\end{multline*}
the 4D analog of (\ref{eq:splitCube}), to cancel the gauge-dependence in $V_\text{ring}^{[A]}(\phi_0,T)$.  Therefore, the prescription we propose for obtaining the gauge-independent expression for $\Delta V_\text{ring}$ is
\begin{equation}\label{eq:GIringPres}
\Delta V_\text{ring}^\text{G.I.}(T)=V_\text{ring}^{[B]}(\phi_0(T),T)-V_\text{ring}^{[A]}(\phi_0,T)\,,
\end{equation}
so that the expression for the gauge-independent one-loop ring-improved effective potential at its minima is
\begin{equation}
\label{eq:vring3}
V_\text{eff}(\phi_\text{min}(T),T)=V_0(\phi_0)+\hbar V_1(\phi_0,T)+\Delta V_\text{ring}^\text{G.I.}(T)\,.
\end{equation}

As it stands, the construction of $\Delta V_\text{ring}^\text{G.I.}$ is somewhat \emph{ad hoc} as we have not yet rigorously demonstrated that (\ref{eq:GIringPres}) is the result of re-summing the $g^2T^2/m^2$ terms in (\ref{Vcurve}) to all orders in $\hbar$.  Instead, using a simple $\phi^4$ theory, we illustrate that our prescription correctly gives the hard thermal loop contribution to the $\mathcal{O}(\hbar^2)$ term of the series in (\ref{Vcurve}).

Starting with the tree-level potential of $\phi^4$ theory,
\begin{equation}
V_0(\phi)=-\frac{\mu^2}{2}\phi^2+\frac{\lambda}{4}\phi^4\,,
\end{equation}
we find that the tree-level minimum $\phi_0$ and the field-dependent mass $m^2(\phi)$ are given by
\begin{equation}\label{eq:4dtreemin}
\phi_0=\sqrt{\mu^2/\lambda}\qquad m^2(\phi)=\frac{\partial^2 V_0}{\partial \phi^2}=-\mu^2+3\lambda\phi^2\,.
\end{equation}
Following the procedure outlined in section \ref{sec:GaugeProblem}, a little work shows that the one-loop finite temperature effective potential (\ref{finiteTeffPot}) is given by
\begin{align}\nonumber
V_1(\phi,T)&=V_\text{CW}(\phi)+\frac{T^4}{2\pi^2}J_B\big(m^2(\phi)/T^2\big)\\
&=-\frac{\pi^2}{90}T^4+\frac{T^2}{24}m^2(\phi)-\frac{T}{12\pi}m^2(\phi)^{3/2}+\ldots\,,
\end{align}
from which we derive the one-loop correction to the VEV (\ref{eq:condensate}),
\begin{equation}\label{eq:4dloopmin}
\phi_1(T)=-\frac{\sqrt{\lambda}}{8\mu}T^2+\frac{3\sqrt{2\lambda}}{8\pi}T+\mathcal{O}(T^0).
\end{equation}

We also find the tree-level potential (\ref{eq:VhighTeff0}) of the high temperature effective theory to be
\begin{equation}
V_0^\text{high-$T$}(\phi,T)=\frac{1}{2}\big(-\mu^2+\Sigma(T)\big)\phi^2+\frac{\lambda}{4}\phi^4\,,
\end{equation}
where $\Sigma(T)=\hbar \lambda T^2/4$ is the thermal mass.  The tree-level minimum of this potential is given by
\begin{equation}\label{eq:highTtreemin}
\phi_0(T)=\sqrt{\big(\mu^2-\Sigma(T)\big)/\lambda}\,.
\end{equation}
Then, the result of re-summing the ring diagrams Fig. \ref{fig:Ring1} gives (\ref{eq:ringV})
\begin{multline}\label{eq:ringsumresultphi4}
\Delta V_\text{ring}(\phi,T)=V_\text{ring}^{[B]}(\phi,T)-V_\text{ring}^{[A]}(\phi,T)\\
\hspace{-2cm}=-\hbar\frac{T}{12\pi}\left[\left(m^2(\phi)+\Sigma(T) \right)^{3/2}-\big(m^2(\phi)\big)^{3/2}\right]\,.
\end{multline}

The prescription to construct the gauge-independent potential is to evaluate the first term term in (\ref{eq:ringsumresultphi4}) at $\phi_0(T)$ given by (\ref{eq:highTtreemin}) and the second term at $\phi_0$ given by (\ref{eq:4dtreemin}).  The result is
\begin{align}\nonumber
\Delta V_\text{ring}^\text{G.I.}(\phi,T)&=-\hbar\frac{T}{12\pi}\left[\left(2\mu^2-\hbar\frac{\lambda T^2}{2}\right)^{3/2}-(2\mu^2)^{3/2}\right]\\
\label{eq:vringgiresult}
&=\hbar^2\frac{\lambda\mu}{8\sqrt{2}\pi}T^3+\mathcal{O}(\hbar^2)\,.
\end{align}
We now show that this $\mathcal{O}(\hbar^2T^3)$ term is precisely what is contained in the $\mathcal{O}(\hbar^2)$ term of (\ref{Vcurve})
\begin{equation}\label{eq:Vcurve2}
\mathcal{O}(\hbar^2):\enspace\big[V_2(\phi_0,T)-\textstyle\frac{1}{2}\phi_1^2(T) \frac{\partial^2 V_0}{\partial\phi^2}|_{\phi_0}\big]\,.
\end{equation}
An explicit calculation of the two-loop bubble graph of the type in Fig. \ref{fig:Ring1} contributing to $V_2(\phi_0)$ yields
\begin{equation}\label{eq:bubbgraph}
V_2(\phi_0,T)|_\text{bubb.}=
\parbox{1.5cm}{
\begin{fmffile}{Ring1}
\begin{fmfgraph*}(45,45)
  \fmfset{arrow_len}{3mm}
  \fmfpen{0.6pt}
  \fmfbottom{i1}
  \fmftop{i2}
  \fmf{phantom,tension=7}{i1,v1}
  \fmf{phantom,tension=7}{i2,v3}
  \fmf{dashes,tension=0.5,left}{v1,v2,v1}
  \fmf{plain,tension=1,left}{v2,v3,v2}
\end{fmfgraph*}
\end{fmffile}}
=\frac{-\lambda\mu}{16\sqrt{2}\pi}T^3\,.
\end{equation}
Then, substituting (\ref{eq:4dtreemin}), (\ref{eq:4dloopmin}) and (\ref{eq:bubbgraph}) into (\ref{eq:Vcurve2}) we find
\begin{equation}
(\text{eqn. \ref{eq:Vcurve2}})=\ldots+\frac{\lambda\mu}{8\sqrt{2}\pi}T^3+\mathcal{O}(T^4)\,,
\end{equation}
that the $\mathcal{O}(\hbar^2T^3)$ term is in agreement with (\ref{eq:vringgiresult}).

We have carried out a similar analysis, drawing on the results in Ref.~\cite{Fodor:1994bs}, to show that the prescription given in (\ref{eq:GIringPres}) yields the $\mathcal{O}(\hbar^2 T^3)$ in the SM. We remind the reader that while this result is not a rigorous demonstration of the validity of this prescription, it does indicate that the gauge-independent quantity $\Delta V_\text{ring}^\text{G.I.}$ gives an all-orders estimate of the effects of plasma screening which correctly reproduces the corresponding $\mathcal{O}(\hbar^2T^3)$ contribution arising from the full potential. In section \ref{sec:numerics}, we will study the numerical impact of including $\Delta V_\text{ring}^\text{G.I.}$ when determining $T_C$. 

\subsubsection{Two-loop order}

In addition to investigating the all-orders re-summation of the plasma damping corrections, one may also inquire about the remaining higher-order corrections. To that end, we consider the explicit $\mathcal{O}(\hbar^2)$ terms in Eq.~(\ref{Vcurve}). We first observe that this contribution contains the term
\be
\label{eq:backreact}
-\frac{1}{2}\phi_1^2\frac{\partial^2 V_0}{\partial\phi^2}|_{\phi_0}
\ee
which is manifestly negative due to the concavity of $V_0$ at the minimum. By itself, this negative contribution will lead to an increase in $T_C$, and is therefore suggestive of the importance of remaining $\mathcal{O}(\hbar^2)$ contributions. Of course, gauge-invariance requires inclusion of the explicit two loop term $V_2 (\phi_0,T)$. At present, we are not aware of any computation of $V_2$ in an arbitrary gauge in the full theory\footnote{The corresponding computation in the dimensionally-reduced, high-$T$ effective theory has been carried out in Ref.~\cite{Laine:1994zq}; see above.}  and are unable to rigorously demonstrate the gauge-invariance determination of the $\mathcal{O}(\hbar^2)$ contribution. 

On the other hand, the authors of Ref.~\cite{Fodor:1994bs} have computed this contribution in the full theory for the SM using the Landau gauge. Thus, use of their results in combination with a computation of (\ref{eq:backreact}) in the SM with same gauge should yield a gauge-independent result. Unfortunately, only expressions for $V_2(\phi,T)$ in the high-$T$ expansion are available from  Ref.~\cite{Fodor:1994bs}, so it is not possible in the present study to provide a gauge-independent determination of the complete $\mathcal{O}(\hbar^2)$ contributions. We again defer a full two-loop computation to future study. Nevertheless, to the extent that the high-$T$ approximation to  $V_2(\phi,T)$ provides a numerically realistic estimate of the full result, one may use the former -- together with our computation of (\ref{eq:backreact}) to obtain an indication of complete $\mathcal{O}(\hbar^2)$ impact. In Section \ref{sec:numerics} we show that the trend from the $\mathcal{O}(\hbar^2)$  terms is to raise $T_C$, though the precise value should be taken with a grain of salt. 

\section{Gauge-independent sphaleron energy scale}
\label{sec:sphaleron}
The second factor appearing in the BNPC (\ref{eq:exp}) is the sphaleron scale, commonly taken to be the scalar field $\phi_\text{min}(T_C)$ minimizing the effective potential.  As emphasized above, Nielsen's identity for $V_\text{eff}(\phi,T)$ implies that the background field is an inherently gauge-dependent quantity.  Therefore, we abandon $\phi_\text{min}(T_C)$, and we revisit the computation of the sphaleron rate in search for a gauge-independent scale. In doing so, we recapitulate the work of Refs. \cite{Carson:1990jm,Arnold:1987mh,Carson:1989rf}, wherein an appropriately gauge-independent formulation of a perturbative sphaleron rate computation is developed. Along the way, we point out what we believe has motivated a departure from this framework,  leading to the introduction of additional gauge-dependence via the appearance of $\phi_\text{min}(T_C)$ in the BNPC. We then advocate a return to the gauge-independent treatment.

The sphaleron rate $\Gamma_\text{sph}$ is proportional to the free energy of the sphaleron gas $F_\text{s.g.}$  . For a dilute gas, it is related to the effective action for one sphaleron $\Gamma_\text{eff}[\phi^\text{sph}]$ normalized by the energy density of the electroweak broken phase $\Gamma_\text{eff}[\phi^{EW}]$ \cite{Arnold:1987mh}:
\begin{align}\nonumber
\Gamma_\text{sph}&=\frac{\omega_-}{\pi T}\,\text{Im }F_\text{s.g.}=\frac{\omega_-}{\pi}\text{Im }\frac{Z_\text{sph}}{Z_0}\\
\label{eq:sphRate1}&=\frac{\omega_-}{\pi}\text{ Im }e^{-\Gamma_\text{eff}[\phi^\text{sph}]-\Gamma_\text{eff}[\phi^\text{EW}]}\,,
\end{align}
where $\omega_-$ is the frequency associated with the unstable mode of the sphaleron. Since the sphaleron represents a saddle point of the energy functional, the rate is expected to be gauge-independent on account of Nielsen's identity.  In order to ensure gauge-independence within the perturbative context, the effective action must be computed in manner that is consistent with the $\hbar$-expansion in analogy with (\ref{Vcurve}):
\begin{equation}\label{eq:hbarEffAct}
\Gamma_\text{eff}[\phi^\text{sph};T]=S[\phi_0^\text{sph}]+\hbar\Gamma_1[\phi^\text{sph}_0]+\mathcal{O}(\hbar^2)\,.
\end{equation}
Here, $S[\phi^\text{sph}_0]$ is the tree-level action and $\Gamma_1[\phi^\text{sph}_0]$ is the one-loop fluctuation determinant, both evaluated around the tree-level sphaleron configuration $\phi^\text{sph}_0$.  

\subsection{Dimensional reduction}
For any extension of the SM, the electroweak sphaleron lives in an SU(2) subgroup of the full gauge group. Hence,  we shall simplify our analysis by restricting ourselves to the minimal standard SU(2) model.  The task at hand is to compute the path integral in (\ref{genFunctional}) and (\ref{effAction}) around the sphaleron background.  In general, including finite temperature loop contributions to $\Gamma_\text{eff}[\phi^\text{sph}]$ is considerably more challenging than doing so for $V_\mathrm{eff}(\phi,T)$. 
The calculation may be simplified, however, when temperatures are sufficiently high such that the $n>0$ Matsubara modes may safely be integrated out.  As discussed in the previous section, this dimensional reduction procedure is  gauge-dependent, but the Wilson coefficients of the $\mathcal{O}(T^2)$ terms are gauge-independent.  Gauge independence is maintained by retaining just these terms in the effective theory.  

After imposing the temporal-axial gauge $W_{\mu=0}=0$ on the SU(2) gauge fields $W_\mu^a$, the path integral for static field configurations reads
\begin{gather}\label{eq:sphFuncInt}
Z[j]=\int \mathcal{D}H\, \mathcal{D}W\, e^{-S_3[H,W;T]}\\
\label{sphEnergyFunc}
S_3=\frac{1}{T}\int d^3x\Big[|D_i H|^2+\frac{1}{4}W_{ij}^a W_{ij}^a+V(H,T)\Big]\,,
\end{gather}
with
\begin{align}\label{eq:DimRedPot}
\nonumber V(H,T)&=2D(T^2-T_0^2)H^\dag H+\lambda(H^\dag H)^2+\frac{D^2{T}_0^4}{\lambda}\\
&= \lambda(H^\dag H-\bar{v}(T)^2/2)^2\,,
\end{align}
where as before $2DT^2$ is the Higgs thermal mass;  $2D T_0^2=\mu^2$ is the zero-temperature Higgs mass parameter; and  
\begin{equation}\label{giScale}
\bar{v}(T)=v_0\sqrt{1-T^2/T_0^2}\,,
\end{equation}
with $v_0=246\text{ GeV}$ being the tree-level expectation value of the Higgs field.  Notice that $H^\mathsf{T}=(0,\, {\bar v(T)}/\sqrt{2})$ 
minimizes $V(H,T)$ for $T<T_0$.  It is over this action that the path integral is performed to obtain the sphaleron rate.

\subsection{Tree-level sphaleron energy $\Delta E_\text{sph}$}
We search for the tree-level sphaleron solution around which the path integral is to be computed.  The Klinkhammer and Manton\cite{Klinkhamer:1984di} ansatz for the sphaleron, motivated by the electroweak instanton, is\footnote{The minus sign in the ansatz for $W_i^a$ is due to our sign convention of the covariant derivative $D_i=\partial_i+igW_i$.}
\begin{align}
\label{gaugeAnsatz} W_i^a(\mathbf{x})&=-\frac{2}{g}\frac{f(\mathbf{x}^2)}{\mathbf{x}^2}\epsilon^a_{\phantom{a}ij}x_j\\
\label{higgsAnsatz} H(\mathbf{x})&=h(\mathbf{x}^2)\hat{x}_i\sigma_i\left(\begin{array}{c}0\\\bar{v}(T)/\sqrt{2}\end{array}\right)\,,
\end{align}
which we collectively denote as $\phi^\text{sph}_0$.  Here, $f(\mathbf{x}^2)$ and $h(\mathbf{x}^2)$ are the gauge and Higgs field radial profile functions, respectively. They are determined by imposing the stationary conditions on the energy functional and  satisfy the boundary conditions,
\begin{gather}
f(0)=h(0)=0\\
f(\infty)=h(\infty)=1\,.
\end{gather}
Substituting (\ref{gaugeAnsatz}) and (\ref{higgsAnsatz}) into (\ref{sphEnergyFunc}) yields the energy functional in terms of the radial profile functions,
\begin{equation}
\label{eq:sphenergy}
S_3[\phi^\text{sph}_0]\equiv \frac{\Delta E_\text{sph}}{T}=\frac{4\pi \bar{v}(T)}{gT}\, B\bigg(\frac{\lambda}{g^2}\bigg)\,,
\end{equation}
with
\begin{multline}
\label{eq:B}
B\Big(\frac{\lambda}{g^2}\Big)=\int_0^\infty d\bar{r}\Big[4 \Big(\frac{df}{d\bar{r}}\Big)^2+\frac{8}{\bar{r}^2}f^2(1-f)^2\\
+\frac{\bar{r}^2}{2}\Big(\frac{dh}{d\bar{r}}\Big)^2+h^2(1-f)^2+\frac{1}{4}\Big(\frac{\lambda}{g^2}\Big)\bar{r}^2(1-h^2)^2\Big]\,,
\end{multline}
where $\bar{r}=g \bar{v}(T)|\mathbf{x}|$ is a dimensionless radial coordinate.  The stationary condition for the radial profile functions are
\begin{align}
\bar{r}^2\frac{d^2 f}{d\bar{r}^2}&=2f(1-f)(1-2f)-\frac{1}{4}\bar{r}^2 h^2(1-f)\\
\frac{d}{d\bar{r}}\left(\bar{r}^2\frac{dh}{d\bar{r}}\right)&=2h(1-f)^2-\frac{\lambda}{g^2}\bar{r}^2h(1-h^2)
\end{align}
and must be solved numerically.  With the radial profile functions, $f(\bar{r})$ and $h(\bar{r})$ known, the function $B(\lambda/g^2)$ may be numerically evaluated, providing a quantitative result for $S_3[\phi_0^\text{sph}]$.  The dimensionless function $B(\lambda/g^2)$ is an $\mathcal{O}(1)$ weakly varying function of its argument \cite{Klinkhamer:1984di}.  Therefore, the quantity which sets the scale of the sphaleron energy is $\bar{v}(T)$.

Two points about $\bar{v}(T)$ and $T_0$ should be emphasized. First,  because the scale $\bar{v}(T)$ introduced by dimensional reduction minimizes $V(H,T)$, it is tempting to interpret $\bar{v}(T)$ as $\phi_\mathrm{min}(T)$, the value background Higgs field at temperature $T$.  As discussed above, however, they only coincide in the high-$T$ effective theory where the $T\phi^3$ terms essential to bubble nucleation are absent.  Consequently, it is best within the context of the calculation of the sphaleron rate to treat $\bar{v}(T)$ as an auxiliary quantity that sets the scale of the sphaleron energy.  

Similarly, the temperature $T_0$ that appears in the expression (\ref{giScale}) for $\bar{v}(T)$ is not the same as the critical temperature. Rather, $T_0$ is the temperature at which the sphaleron energy (\ref{eq:sphenergy}) vanishes at tree-level in the dimensionally reduced, high-$T$ effective theory.  At sufficiently high temperatures ($T\geq T_0$), the sphaleron rate contains no exponential suppression that is crucial for preservation of the baryon asymmetry.  As the transition becomes more weakly first order $T_C$ approaches $T_0$ from below, leading to greater washout of the baryon asymmetry. 

\subsection{One-loop fluctuation determinant} 
The calculation of one-loop effective action $\Gamma_\mathrm{eff}(\phi^\text{sph})$ is performed following the background field method, in which the background field is the tree-level sphaleron configuration $\phi^\text{sph}_0$ determined above.  The functional integration is performed in the gaussian approximation, formally giving ratios of functional determinants
\begin{multline}
\Gamma_\text{eff}[\phi^\text{sph}]=S_3[\phi^\text{sph}_0]+\frac{\hbar}{2}\ln\bigg(\frac{\det \mathcal{O}_\text{bos}(\phi^\text{sph}_0)}{\det \mathcal{O}_\text{bos}(\phi^\text{EW})}\bigg)\\
\label{eq:fluctdet}
-\hbar \ln \bigg(\frac{\det \mathcal{O}_\text{FP}(\phi^\text{sph}_0)}{\det \mathcal{O}_\text{FP}(\phi^\text{EW})}\bigg)\,,
\end{multline}
where $\mathcal{O}_\text{bos}$ and $\mathcal{O}_\text{FP}$ are the bosonic and ghost fluctuation operators, whose eigenvalues are proportional to $(\bar{v}/gT)^{1/2}$ as a result of coordinate rescaling.

A precise calculation of these determinants is challenging, but the essential structure is as follows.  The sphaleron background breaks three rotational and three translational symmetries of the system, giving rise to six zero modes in the spectrum of $\mathcal{O}_\text{bos}(\phi_0^\text{sph})$.  The integration over these modes give rise to volume prefactors $(\mathcal{N}\mathcal{V})$ and six factors of the proportionality constant $(\bar{v}/gT)^{1/2}$. Here, $\mathcal{N}$ and $\mathcal{V}$ are normalization and dimensionless integration volume factors, respectively. 
Letting $\kappa$ denote the ratio of fluctuation determinants with zero modes removed, the sphaleron rate (\ref{eq:sphRate1}) then takes the form
\begin{equation}\label{eq:sphRate}
\begin{aligned}
\Gamma_\text{sph}&=\frac{\omega_-}{2\pi}(\mathcal{N}\mathcal{V})_\text{tr}(\mathcal{N V})_\text{rot}\bigg(\frac{\bar v}{gT}\bigg)^3\kappa e^{-\Delta E_\text{sph}/T}\\
&=\frac{\omega_-}{2\pi}(\mathcal{N}V)_\text{tr}(\mathcal{N V})_\text{rot}(g\bar{v})^3\bigg(\frac{\bar v}{gT}\bigg)^3\kappa e^{-\Delta E_\text{sph}/T}\,,
\end{aligned}
\end{equation}
where in the second line we have expressed the dimensionless volume factor $\mathcal{V}_\text{tr}$ in terms of the physical volume $V_\text{tr}$ using $\bar{r}=g\bar{v}(T)|\mathbf{x}|$.  The factor of $1/2$ arises from an analytic continuation of the integration over the negative mode.

The technically involved and numerically intensive task of evaluating $\kappa$ for the standard SU(2) model was achieved by Carson, Li, McLerran  and Wang\cite{Carson:1990jm}, and by Baacke and Junker\cite{Baacke:1993aj,Baacke:1994ix}.  To our knowledge, similar perturbative computations do not exist for SM extensions.  Rather, the conventional practice has been to attempt  to approximate the one-loop contribution by replacing the potential $V(H,T)$ in (\ref{eq:DimRedPot}) with the one-loop effective potential $V_\text{eff}(\phi,T)$ in (\ref{finiteTeffPot}) derived from the full theory. In principle, doing so includes some part of the one-loop fluctuation determinants, but clearly neglects contributions involving spatial variations in the scalar and gauge fields.  

Following this practice and repeating the steps leading to (\ref{eq:sphenergy}) gives a similar result but with the gauge-independent scale $\bar{v}(T)$ replaced by the gauge-dependent Higgs field $\phi_\text{min}(T)$, leading to
\begin{equation*}
\Gamma_\text{sph}\rightarrow\frac{\omega_-}{2\pi}(\mathcal{N}V)_\text{tr}(\mathcal{N V})_\text{rot}(g\phi_\text{min})^3\bigg(\frac{\phi_\text{min}}{gT}\bigg)^3\kappa e^{-\Delta E_\text{sph}/T}\,,\\
\end{equation*}
where
\begin{equation*}
\frac{\Delta E_\text{sph}}{T}\rightarrow\frac{4\pi \phi_\text{min}(T)}{gT}\tilde{B}\bigg(\frac{\lambda}{g^2}\bigg)\,.
\end{equation*}
For a given choice of gauge, $\phi_\text{min}(T)$ na\"ively approximates the full background field more closely than does $\bar{v}(T)$, so it may seem plausible that the replacement would lead to a better estimation of the sphaleron rate.  By integrating the rate law using this expression for the sphaleron rate to derive the washout criteria (details in Appendix A), one arrives at the criterion in equation (\ref{eq:exp}).

In view of our discussion of $V_\text{eff}(\phi, T)$ it is immediately apparent that  making the foregoing replacement introduces gauge-dependence into $\Delta E_\text{sph}/T$---and thereby into $\Gamma_\text{sph}$---at any temperature. Moreover, as observed in Ref.~\cite{Strumia:1998nf}  in the context of false vacuum decay, this procedure double counts the $n=0$ Matsubara contributions to the sphaleron rate. Since these degrees of freedom already contribute non-analytic terms to $V_\mathrm{eff}(\phi, T)$ from which $\phi_\text{min}(T)$ is obtained, including them in the fluctuation determinant calculation would count them twice. 

In light of these two problems, we advocate abandoning the \emph{ad hoc} replacement $V(H,T)\rightarrow V_\text{eff}(\phi,T)$ in favor of adhering to the gauge-independent formulation of Ref. \cite{Arnold:1987mh} and using $\bar{v}(T)$ instead of $\phi_\text{min}(T)$ to characterize the sphaleron rate.  When attempting to approximately establish the efficacy of baryon number preservation, one should then use the gauge-independent ratio of $\bar{v}(T_C)/T_C$ with $T_C$ computed using the $\hbar$-expansion as above and ${\bar v}(T)$ obtained from the dimensionally-reduced effective theory in the high-$T$ approximation. Including $\mathcal{O}(\hbar)$ contributions to $\Gamma_\mathrm{sph}$ associated with Matsubara zero mode fluctuations about the sphaleron the requires a careful evaluation of the functional determinant. 

\section{An Illustrative Application: $\bar{v}(T_C)/ T_C$ in the Standard Model}
\label{sec:SM}
As an illustrative application,  we now apply our gauge-independent method to analyze the EWPT at $\mathcal{O}(\hbar)$ in the SM, and estimate the impact of higher order contributions. Since there exist Monte Carlo results for both $T_C$ and $\Gamma_\text{sph}$ in the SM, we may compare the corresponding quantities obtained with the gauge-independent perturbative analysis to assess the reliability of the latter. As a result, we will conclude that inclusion of two-loop contributions to $V_\text{eff}(\phi,T)$ is likely to be essential for obtaining a reasonable value of $T_C$. Later, when discussing the criteria for preservation of the baryon asymmetry (section \ref{sec:BNPC}), we will argue that a reliable perturbative computation of the fluctuation determinant is also likely to be important, though probably less decisive than the evaluation of $T_C$.
We also argue that even when gauge-independence is maintained, a perturbative treatment is at best indicative and that definitive statements about baryon number preservation are likely to require non-perturbative studies.

The two ingredients needed for this analysis are the critical temperature and the sphaleron scale.  While contributions from fermions have been previously omitted, our numerical results include effects from the top quark.  We begin by computing the effective potential for the neutral Higgs field, including the effects of the top quark. In the standard $\text{SU(2)}\times\text{U(1)}$ model, the four real degrees of freedom
\begin{equation}
  H=\frac{1}{\sqrt{2}}\left(\begin{array}{c}\Phi_1+i\Phi_2\\\Phi_3+i\Phi_4\end{array}\right)
\end{equation} 
are placed into a vector $\Phi_i=(\Phi_1,\,\Phi_2,\,\Phi_3,\,\Phi_4)$\,.  Explicitly, the generators $T^a$ in the real representation ($-i$ factored out) are
\begin{gather*}
\begin{aligned}
T^1&=\textstyle\frac{1}{2}\left(\begin{smallmatrix}&&&-1\\&&1&\\&-1&&\\1&&&\end{smallmatrix}\right)\\
T^3&=\textstyle\frac{1}{2}\left(\begin{smallmatrix}&-1&&\\1&&&\\&&&1\\&&-1&\end{smallmatrix}\right)
\end{aligned}\qquad
\begin{aligned}
T^2&=\textstyle\frac{1}{2}\left(\begin{smallmatrix}&&1&\\&&&1\\-1&&&\\&-1&&\end{smallmatrix}\right)\\
T^4&=\textstyle\frac{1}{2}\left(\begin{smallmatrix}&-1&&\\1&&&\\&&&-1\\&&1&\end{smallmatrix}\right)\,,
\end{aligned}
\end{gather*}
where the $\{T^1,\,T^2,\,T^3\}$ are generators of isospin and $T^4$ is hypercharge.  The Euclidean Lagrangian takes the form 
\begin{gather}
\mathcal{L}_\text{E}=\frac{1}{2}(D_\mu \Phi)_i (D_\mu\Phi)_i +V(\Phi)\,,\\
\text{with }V(\Phi)=-\frac{1}{2}\mu^2\Phi_i\Phi_i+\frac{1}{4}\lambda(\Phi_i\Phi_i)^2\,,
\end{gather}
where $D_\mu=\partial_\mu+gT^aW_\mu^a$, and $W^a_\mu=\{W^1_\mu,W^2_\mu,W^3_\mu,B_\mu\}$.  We shift the scalar fields following the background field method,
\begin{gather}
\Phi_i(x)\rightarrow \phi_i(x)+\bar\phi\,,\\
\bar\phi=(0,\,0,\,h,\,0)\,,
\end{gather}
and we immediately obtain the tree-level potential in terms of the classical field $h$,
\begin{equation}
V_0(h)=-\frac{1}{2}\mu^2 h^2+\frac{1}{4}\lambda h^4\,.
\end{equation}
We record here the extrema of the tree-level potential:
\begin{gather}\label{eq:SMphases}
h_0^{(1)}=0\,, \qquad h_0^{(2)}=\pm\sqrt{\mu^2/\lambda}=\pm246\text{ GeV}.
\end{gather}  
Next we impose the gauge-fixing condition, $\partial^\mu W_\mu^a+\xi \phi_i(gT^a\bar\phi)_i=0$ on the quantum fields.  Retaining terms quadratic in fluctuations, we read off the field-dependent mass matrices necessary to construct the effective potential:
\begin{align}\label{SMscalarMtx}
M_{ij}^2(h)&=\begin{pmatrix}-\mu^2+\lambda h^2&&&\\&\hspace{-7mm}-\mu^2+\lambda h^2&&\\&&\hspace{-7mm}-\mu^2+3\lambda h^2&\\&&&\hspace{-7mm}-\mu^2+\lambda h^2\end{pmatrix}\\
\label{SMgoldstoneMtx}
m_A^2(h)_{ij}&=\frac{1}{4}\begin{pmatrix}g^2&&&\\&g^2&&\\&&0&\\&&&(g'^2+g^2)\end{pmatrix}h^2\\
\label{SMgaugeMtx}
m_A^2(h)^{ab}&=\frac{1}{4}\begin{pmatrix}g^2&&&\\&g^2&&\\&&g^2&-gg'\\&&-gg'&g'^2\end{pmatrix}h^2\,.
\end{align}
After abbreviating their eigenvalues
\begin{gather}\label{eq:SMfieldDepmasses}
\begin{aligned}
m_H^2(h)&=-\mu^2+3\lambda h^2\,,\\
m_G^2(h)&=-\mu^2+\lambda h^2\,,\\
m_\gamma^2(h)&=0\,,
\end{aligned}\enspace
\begin{aligned}
m_Z^2(h)&=\textstyle\frac{1}{4}(g^2+g'^2)h^2\,,\\
m_W^2(h)&=\textstyle\frac{1}{4}g^2h^2\,,\\
\phantom{m_Z^2}
\end{aligned}
\end{gather}
we write down the 1-loop finite-temperature effective potential (\ref{finiteTeffPot})
\begin{equation}
V_\text{eff}(h)=V_0(h)+\hbar \big[V_1^{T=0}(h)+V^{T\neq0}_1(h,T)\big]\,,
\end{equation}
where explicit expressions in the $\overline{\text{MS}}$ renormalization scheme are provided in Appendix  \ref{app:VeffSM}.
To determine the critical temperature, we use (\ref{Vcurve}) to follow the temperature-evolution of the free energy of the symmetric $h^{(1)}_0$ and broken $h^{(2)}_0$ phases.  Then the $\mathcal{O}(\hbar)$ degeneracy condition defining the critical temperature of the phase transition reads
\begin{multline}\label{eq:SMdegencond}
V_0(h_0^{(1)})+\hbar V_1(h_0^{(1)},T_C)=V_0(h_0^{(2)})+\hbar V_1(h_0^{(2)},T_C)\,.
\end{multline}

Before we proceed with numerical results, we explicitly show that both sides of (\ref{eq:SMdegencond}) are gauge-independent, referring to (\ref{eq:SMTindepVeff}) and (\ref{eq:SMTdepVeff}).  The statement is trivial for the LHS, since the one-loop potentials $V_1^{T=0}(h)$ and $V_1^{T\neq0}(h,T)$ are evaluated at the origin $h_0^{(1)}=0$.  At that point, $m_W^2$ and $m_Z^2$ both vanish and all gauge-dependent terms that multiply them disappear.  In the RHS, the potential is evaluated at the tree-level broken phase minimum $h_0^{(1)}=\sqrt{\mu^2/\lambda}$.  In this case $m_G^2=0$, and the gauge-dependent terms cancel.

Let us illustrate how theorem 1 in Appendix \ref{app:VGM} operates for the matrices in eqns. (\ref{SMscalarMtx}) and (\ref{SMgoldstoneMtx}).  Notice that the sum $M_{ij}^2(h)+\xi m_A^2(h)_{ij}$ that goes into the computation of the effective potential (\ref{TeffPotTrLog}), for general values of the Higgs background field $h$, has mixed dependence on the gauge-parameter.  This blocks us from being able to split logarithms as in (\ref{TrLogSplit}).  However when evaluated at the tree-level minimum $h_0^{(2)}=\pm\sqrt{\mu^2/\lambda}$, we find 
\begin{align*}
M_{ij}^2(h_0^{(2)})&=
\mathrm{diag}\left(
0,\,0,\, 2\mu^2,\, 0\right)\\
m_A^2(h_0^{(2)})_{ij}&=\frac{\mu^2}{4\lambda}\mathrm{diag}\left(
g^2,\, g^2,\, 0,\, (g'^2+g^2)\right)\,,
\end{align*}
that these matrices are (trivially) simultaneously diagonalizable, and more importantly, their non-zero eigenvalues reside in distinct subspaces.  This property enables us to split the logarithm as in (\ref{TrLogSplit}).

We can also verify theorem 2 for the matrices in Eqs. (\ref{SMgoldstoneMtx}) and (\ref{SMgaugeMtx}) by diagonalizing the gauge boson mass matrix $m_A^2(h^{(2)}_0)^{ab}$ via a rotation through the weak mixing angle $\theta_W$:
\begin{equation*}
R^{ab} m_A^2(h)^{bc}(R^\mathsf{T})^{cd}= \frac{1}{4}\begin{pmatrix}g^2&&&\\&g^2&&\\&&0&\\&&&(g'^2+g^2)\end{pmatrix}h^2\,,
\end{equation*}
where
\begin{equation*}
R^{ab}=\begin{pmatrix}1\\ &1&&\\ &&\cos\theta_W&-\sin\theta_W\\ &&\sin\theta_W&\cos\theta_W\end{pmatrix},\qquad \tan\theta_W=g'/g\,.
\end{equation*}
Therefore, the matrices $m_A^2(h^{(2)}_0)^{ab}$ and $m_A^2(h_0^{(2)})_{ij}$ have the same non-zero eigenvalues. Thus, when carrying out the sum over eigenvalues in (\ref{CWeffPot}) and (\ref{finiteTeffPot}), the gauge-dependent terms cancel.

Finally, as outlined in section \ref{sec:sphaleron}, the gauge-independent sphaleron scale $\bar{v}(T)$ is derived from the high-temperature effective theory.  In the standard model, the sphaleron scale is given by (\ref{giScale}) and (\ref{eq:coeffdef}).

\subsection{Numerical results}
\label{sec:numerics}
\begin{figure}
\includegraphics[scale=1.0,width=7cm]{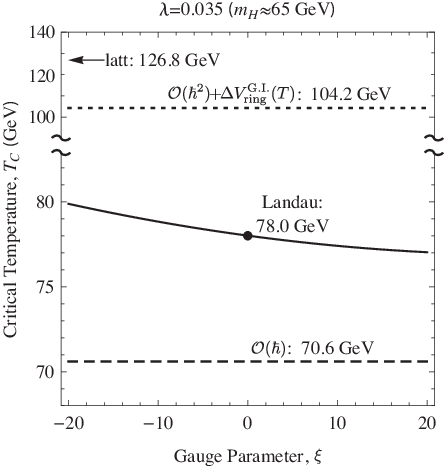}
\caption{A comparison of the critical temperatures as computed using the following methods: the standard method (solid line); the gauge-independent methods described in the text, derived from the full theory (dashed lines); and by performing lattice simulations (arrow).  Note that the lattice result is much higher than the perturbative estimations and is displayed on a separate scale.}\label{TdepSM}
\end{figure}

In Fig. \ref{TdepSM}, we compare critical temperatures derived following our gauge-independent method with the one derived using the conventional procedure as a function of the gauge parameter $\xi$.  In this example, we have chosen $\lambda=0.035$ corresponding to a low Higgs mass of approximately $65\text{ GeV}$ -- close to the phase transition end-point according to lattice studies.  

In this plot, we display the gauge-independent results for $T_C$ at two levels of approximation: at $\mathcal{O}(\hbar)$ using (\ref{eq:SMdegencond}), and an estimate at $\mathcal{O}(\hbar^2)$ with ring re-sum (\ref{eq:GIringPres}) using
\begin{equation}
V_\text{eff}^{\mathcal{O}(\hbar^2)}(h^{(1)},T_C)=V_\text{eff}^{\mathcal{O}(\hbar^2)}(h^{(2)},T_C)\,,
\end{equation}
where
\begin{multline}\label{eq:vring5}
V_\text{eff}^{\mathcal{O}(\hbar^2)}(h,T)=V_0(h_0) +\hbar V_1(h_0) +\Delta V_\text{ring}^\text{G.I.}(T)\\
+\hbar^2\big[V_2(h_0,T)-\textstyle\frac{1}{2}h_1^2(T) \frac{\partial^2 V_0}{\partial h^2}|_{h_0}\big]-\mathcal{O}(\hbar^2T^3)\,.
\end{multline}
The $\mathcal{O}(\hbar^2T^3)$ subtraction represents a careful removal of that term so as to not double-count the contribution in $\Delta V_\text{ring}^\text{G.I.}(T)$.  At present we are able to provide only a rough estimate of the $\mathcal{O}(\hbar^2)$ contribution since the expression for $V_2(h,T)$ in Ref.~\cite{Fodor:1994bs} is given in the high-$T$ limit only.  Nonetheless, we expect this estimate to provide an indication of magnitude of effects associated with higher order contributions.
 
Also included are lattice results, that yield a critical temperature of $126.8\text{ GeV}$, independent of $\xi$ by construction. Our estimate of the higher-order contributions included in (\ref{eq:vring5}) leads to a substantially larger value of $T_C$, suggesting that the difference between the non-perturbative and $\mathcal{O}(\hbar)$ perturbative results arises in part from the omission of higher-order contributions. In addition, we note that the precise definition of $T_C$ as obtained from the lattice studies differs from the one we have employed here as well as in other perturbative analyses (For a discussion of the lattice determinations, see, {\em e.g.}, Refs.~\cite{Kajantie:1996mn,Aoki:1999fi,Moore:1998swa}). We speculate that part of the difference between the lattice and perturbative results may also be due to this difference in definition.

\begin{figure}
\includegraphics[scale=1.0,width=7cm]{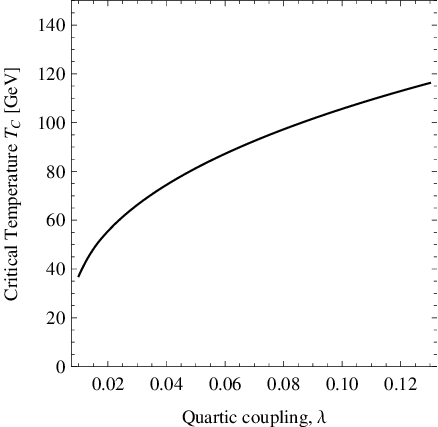}
\caption{Gauge-independent one-loop critical temperature as a function of the SM Higgs quartic self-coupling.}\label{TClambda}
\end{figure}

While the gauge-independent perturbative estimation of $T_C$ falls below the lattice value, it is interesting that the dependence on the relevant couplings follows the trend observed in non-perturbative studies. To illustrate, we plot the one-loop $T_C$ as a function of the Higgs quartic self-coupling $\lambda$ in Fig. \ref{TClambda}. We observe that increasing $\lambda$ increases $T_C$ in agreement with our qualitative expectations in (\ref{eq:giTrend}). As we discuss shortly, this trend implies that the efficiency of sphaleron-induced baryon number washout increases with $\lambda$ and, thus, with the value of the Higgs boson mass. This trend is also observed in non-perturbative studies as well as in earlier gauge-dependent perturbative analyses. 

\begin{figure}
\includegraphics[scale=1,width=7cm]{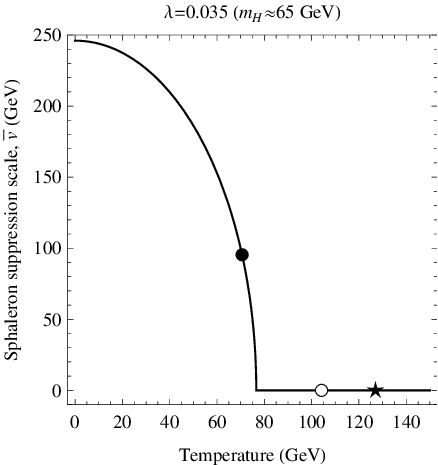}
\caption{Gauge-independent sphaleron scale, $\bar{v}(T)$ in Eq. (\ref{giScale}) as a function of temperature.   Also shown are the gauge-independent $\mathcal{O}(\hbar)$ (solid circle) and approximate $\mathcal{O}(\hbar^2)$ (open circle) estimates of $T_C$ derived from the full theory, and the lattice result (star). }\label{vbarSM}
\end{figure}
We now turn our attention to the sphaleron scale, ${\bar v}(T)$, which we plot in Fig. \ref{vbarSM}.  We observe that in the vicinity of the $T_C$ obtained at $\mathcal{O}(\hbar)$ in the full theory, ${\bar v}(T)$ drops rapidly to zero. This behavior makes the perturbative estimate of the sphaleron rate at the critical temperature highly sensitive to small changes in $T_C$.  Therefore, statements about the efficacy of baryon number preservation susceptible to large uncertainties.  To illustrate, we first consider the value of this scale at the one-loop $T_C$ in the full theory,  $95.4\text{ GeV}$.  We then obtain for the ratio
\begin{equation}
\frac{\bar{v}(T_C)}{T_C}\bigg|_{\mathcal{O}(\hbar)}=1.35\,,
\end{equation}
implying that, at this level of approximation, sphaleron processes may be sufficiently quenched in the standard model to preserve the baryon asymmetry (see Section \ref{sec:BNPC}). This conclusion stands in stark contrast to the conclusion one would reach from the results of Monte Carlo studies, which give a much larger $T_C$---well above $T_0=76.6\text{ GeV}$ at which ${\bar v}(T)$ vanishes. On the other hand, if we take the estimate of $T_C$ at $\mathcal{O}(\hbar^2)$ using (\ref{eq:vring5}), we observe that it falls above $T_0$, suggesting unsuppressed sphaleron transitions that would completely erase the baryon asymmetry. 

In light of this strong sensitivity to higher-order contributions in the vicinity of $T_0$, one should evaluate the degree to which a perturbative analysis is indicative. To gain insight, we again turn to results of Monte Carlo simulations carried out in Ref.~\cite{Moore:1998swa}, which found that for a range of values for $\lambda$ that includes the choice used here, the sphaleron rate $\Gamma_\text{sph}$ would be sufficiently quenched so as to preserve an initial baryon asymmetry, even though the $\mathcal{O}(\hbar)$ gauge-independent value of $T_C$  falls well below the result of Monte Carlo studies. This situation suggests that higher order contributions to $\Gamma_\text{sph}$ may compensate for the larger value of $T_C$, resulting in stronger quenching of the sphaleron transitions at a given temperature than one might infer from perturbation theory. 

It appears, then, that one should treat with caution any conclusions drawn from perturbation theory about the degree of baryon washout for critical temperatures in the vicinity of $T_0$, even when gauge-independence is fully maintained. We believe, nevertheless, that perturbation theory retains some utility in providing guidance as to the relevant regions of parameter space where baryon number preservation is likely to be more or less effective. To illustrate this point, we consider two examples. 

\begin{figure}
\includegraphics[scale=1,width=7cm]{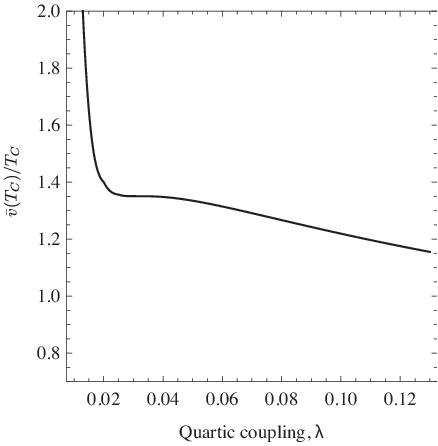}
\caption{Ratio of gauge-independent sphaleron scale and critical temperature as a function of the SM Higgs quartic self-coupling. }\label{vbarTC}
\end{figure}

First, in Fig. \ref{vbarTC} we plot the ratio ${\bar v}(T_C)/T_C$ as a function of $\lambda$. As expected from Monte Carlo studies, this ratio decreases with $\lambda$, implying an increase in the sphaleron rate with $m_H$. The precise value of $\lambda$ at which baryon number preservation is too weak to be phenomenologically viable is uncertain -- a point we elaborate in Section \ref{sec:BNPC}. Nevertheless, the trend is indicative and suggests that any extension of the SM scalar sector that reduces the effective quartic coupling at $T\sim T_C$ is likely to enhance baryon number preservation. As discussed in Ref.~\cite{Profumo:2007wc}, for example, the inclusion of new tree-level operators involving singlet degrees of freedom may lead to just such a weakening of $\lambda_\mathrm{eff}(T_C)$ while respecting the LEP lower bound on the SM-like Higgs boson mass. 

As a second illustration, we consider the MSSM. It has been thought for some time that the presence of a light right-handed top squark is needed in order to preserve a sufficiently large initial baryon asymmetry (see Ref.~\cite{Quiros:1999jp} and references therein). In the conventional, gauge-dependent analysis, this conclusion can be seen by considering the top squark contribution to the ring re-sum
\begin{equation}\label{eq:SUSYthermal}
-\hbar\frac{T}{12\pi}\Big[\big(m^2_{\tilde t}+y_{\tilde t}^2 \phi^2+\Sigma_{\tilde t}(T)\big)^{3/2}-\big(m^2_{\tilde t}+y_{\tilde t}^2 \phi^2\big)^{3/2}\Big]\,,
\end{equation}
where the second term cancels against the corresponding term in the high-$T$ expansion of the potential. Then, the stop soft mass parameter $m_{\tilde t}^2$ must be negative with a magnitude roughly needed to cancel the stop Debye mass contribution $\Sigma_{\tilde t}(T)$ in the first term. In this region of parameter space, the stop contribution to the ring potential leads to an approximate $T\phi^3$ term in the high-$T$ effective potential needed for a first order EWPT. Since the stop Yukawa coupling is $\mathcal{O}(1)$, this contribution can be particularly pronounced. Phenomenologically, the soft mass parameter $m_{\tilde Q_3}^2$ for the left-handed third generation squark doublet must be positive and relatively large in order to generate a phenomenologically allowed mass for the lightest Higgs, leaving a tachyonic right-handed stop as the only alternative. 

\begin{figure}
\parbox{3cm}{
\includegraphics[scale=1.0,width=4cm]{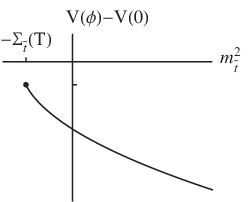}\\ \textbf{a}}
\qquad\qquad
\parbox{3cm}{
\includegraphics[scale=1.0,width=4cm]{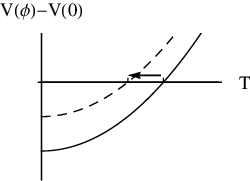}\\ \textbf{b}}
\caption{\textbf{a.}\enspace Behavior of (\ref{eq:vminshift}) as a function of stop mass parameter $m^2_{\tilde t}$.  \textbf{b.}\enspace Broken phase free energy curve (relative to the symmetric phase) as a function of temperature $T$.  A rise in the free energy curve (from solid to dotted line) results in a drop of $T_C$ (indicated by the arrow).}\label{fig:ringSketch}
\end{figure}

Since we have argued that the foregoing analysis is gauge-dependent, one may rightly question whether this scenario remains valid when a gauge-independent perturbative treatment is followed. We believe that it does. We reconsider (\ref{eq:SUSYthermal}) and apply the prescription (\ref{eq:GIringPres}) to construct the ring-improved gauge independent potential.  We evaluate the resulting contribution to the difference between the energies of the electroweak symmetric and broken minima giving
\begin{multline}\label{eq:vminshift}
V_\text{eff}(\phi_\text{min},T)-V_\text{eff}(0,T)\sim\\
-\hbar\frac{T}{12\pi}\Big[\big(m^2_{\tilde t}+y_{\tilde t}^2 \phi^2+\Sigma_{\tilde t}(T)\big)^{3/2}
-\big(m^2_{\tilde t}+\Sigma_{\tilde t}(T)\big)^{3/2}\Big]\,.
\end{multline}
Referring to Fig. \ref{fig:ringSketch}a, for any $m_{\tilde t}^2>-\Sigma_{\tilde t}(T)$ the difference is negative.  But, as $m_{\tilde t}^2$ is lowered closer to $-\Sigma_{\tilde t}(T)$, the difference in free energy rises; \emph{i.e.} the free energy curve of the broken phase is pushed up relative to that of the symmetric phase.  As a result $T_C$ decreases (Fig. \ref{fig:ringSketch}b). This drop in $T_C$ with an increasingly negative soft mass-squared is, in fact, observed in Monte Carlo studies (see, {\em e.g.}, Fig. 9 of Ref. \cite{Laine:1998qk}).  Since ${\bar v}(T_C)/T_C$ correspondingly increases, the baryon asymmetry is more efficiently preserved.

\section{Baryon number preservation criterion}
\label{sec:BNPC}
In previous sections, we have discussed at length the issue of gauge-dependence and its impact on the LHS of the bound in Eq.~(\ref{eq:exp}).  In this section, we revisit the assumptions used to obtain a numeric value appearing on the RHS of the bound.  In Appendix \ref{sec:washout}, we have collected and somewhat re-organized previously-employed results for the BNPC in a manner that makes various sources of uncertainty more apparent. The approximate BNPC, given in (\ref{eq:washout}) is
\begin{multline}
\frac{4\pi B}{g}\frac{\bar{v}(T_C)}{T_C}-6\ln\frac{\bar{v}(T_C)}{T_C} >\\
 -\ln X-\ln\Big(\frac{\Delta t_\text{EW}}{t_H}\Big)+\ln \mathcal{Z}+\hbar\ln\kappa\,.
\end{multline}
Here, $X=-\ln S$ characterizes the dilution of an initial baryon asymmetry during the transition as per (\ref{eq:washoutfactor}), $\Delta t_\mathrm{EW}/t_H$ gives the duration of the transition in units of the Hubble time $t_H$, and $\kappa$ is the ratio of fluctuation determinants in (\ref{eq:fluctdet}) after separating out the zero-mode contributions.  The factor $\mathcal{Z}$ collects a number of remaining terms that enter into the perturbative rate calculation:
\begin{equation}
\mathcal{Z} = \left(\frac{13 n_f}{2}\right)\ \mathcal{N}_\mathrm{tr}\ \left(\mathcal{NV}\right)_\mathrm{rot}\left(\frac{\omega_{-}t_H}{\pi}\right)\,,
\end{equation}
where, $\mathcal{N}_\mathrm{tr}$ and $\mathcal{N}_\mathrm{rot}$ are normalization factors associated with translational and rotational zero mode fluctuations about the sphaleron, $\mathcal{V}_\mathrm{rot}$ is the corresponding rotational volume, $n_f$ is the number of families of fermions, and $\omega_{-}$ is the frequency associated with the unstable mode of the sphaleron.  Taking $t_H\approx (3\times 10^{-2}) M_\mathrm{Pl}/T^2$ with $M_\mathrm{Pl}$ being the Planck mass, and 
\be
\omega_{-}= g{\bar v} \mathcal{F}
\ee
with $\mathcal{F}$ being a function of $T$, ${\bar v}$ and the SM couplings, we arrive at
\begin{multline}
\label{eq:washout2}
\frac{4\pi B}{g}\frac{\bar{v}(T_C)}{T_C}-7\ln\frac{\bar{v}(T_C)}{T_C} >\\
 -\ln X-\ln\Big(\frac{\Delta t_\text{EW}}{t_H}\Big)+\ln \mathcal{QF}+\hbar\ln\kappa\, ,
\end{multline}
with
\be
\label{eq:Qdef}
Q=  (3\times 10^{-2})\times \left(\frac{13 n_f}{2}\right)\ \mathcal{N}_\mathrm{tr}\ \left(\mathcal{NV}\right)_\mathrm{rot} 
\left(\frac{g M_\mathrm{Pl}}{T}\right)\ \ \ .
\ee

From Eqs.~(\ref{eq:washout2}) and (\ref{eq:Qdef}) one may attempt to derive a requirement on the ratio $\bar{v}(T_C)/{T_C}$. The rotational and translational model factors have been computed in Ref.~\cite{Carson:1989rf} as a function of $\lambda/g^2$. These authors find that $\mathcal{N}_\mathrm{tr}\ (\mathcal{NV})_\mathrm{rot}\approx 7000$ and nearly constant for a substantial range in $\lambda/g^2$. The remaining terms in Eqs.~(\ref{eq:washout2}) and (\ref{eq:Qdef}) contain a stronger dependence on the couplings and larger theoretical uncertainties. We have explored the quantitative impact in each case and discuss each below. Before doing so, we adopt a representative set of \lq\lq benchmark" values:
\bea
\nonumber
{\cal F}=1 &\ \ ,\qquad & X=1\\
\label{eq:benchmark1}
\Delta t_\mathrm{EW}/t_H=1 &\ \ ,\qquad & T_C=100\ \mathrm{GeV}\\
\nonumber
\lambda/g^2=1 &\ \ ,\qquad & \kappa=10^{-8}
\eea
For this choice, the inequality in Eq.~(\ref{eq:washout2}) leads to 
\be
\label{eq:washout3}
\frac{\bar{v}(T_C)}{T_C} > 0.73\ \ \ .
\ee
We observe that this requirement is somewhat more relaxed than the one usually quoted in the literature. As we now discuss, however, there exists considerable room for variation in the RHS of Eq.~(\ref{eq:washout3}), which may be nearly a factor of two larger under certain circumstances.
\begin{itemize}
\item[(i)] The function $\mathcal{F}$ depends strongly on which fluctuations dominate the unstable direction, namely, those associated with the Higgs scalar or those associated with the gauge degrees of freedom. The authors of Ref.~\cite{Arnold:1987mh} argue that in the former instance one has $\mathcal{F}=1$ [\lq\lq case (a)"] while in the latter   $\mathcal{F} \approx ({\bar v} /T)^2 \times (1/2\pi)$ [\lq\lq case (b)"].  Our benchmark choice corresponds to case (a). Generally speaking, case (b) leads to a more relaxed requirement on $\bar{v}(T_C)/{T_C}$, reducing the RHS by $\sim 0.1$.
\item[(ii)] The benchmark critical temperature lies between the one-loop, gauge-independent perturbative value and the value obtained from Monte Carlo studies. Increasing it to the lattice value lowers the RHS of Eq.~(\ref{eq:washout3}) five percent or less. 
\item[(iii)] Our choice of $X=1$ corresponds to allowing an initial baryon asymmetry to be diluted by a factor of $1/e$ over the course of the EWPT. In earlier work (see, {\em e.g.}, Ref.~\cite{Quiros:1999jp}), a value of $X\approx 10$ was used, corresponding to a dilution of roughly five orders of magnitude. In this case, the RHS of Eq.~(\ref{eq:washout3}) becomes smaller by ten percent or more. However, recent computations of the initial baryon asymmetry using state-of-the art transport theory imply that such a dilution is likely to be wildly unrealistic (see, {\em e.g.}, Ref.~\cite{Chung:2009qs} and references therein). In the MSSM, for example, one requires a rather narrow window on the relevant CP-violating parameters to obtain an initial baryon asymmetry that is close to the present value \cite{Chung:2009qs,Cirigliano:2009yd}. A more realistic value of $X$ would likely lie in the range 0.01 to 0.1, corresponding to a dilution of one to ten percent. Choosing $X=0.01$ increases the RHS of Eq.~(\ref{eq:washout3}) by order 30\%.
\item[(iv)] According to the numerical work of Ref.~\cite{Carson:1990jm}, the value of the fluctuation determinant ratio $\kappa$ depends strongly on $\lambda/g^2$. For a Higgs mass of $\sim 65$ GeV, one has $\kappa\approx 10^{-8}$, which we have used as our benchmark value. Increasing $m_H$ to the LEP lower bound (or $\lambda/g^2\approx 0.3$), increases $\kappa$ to $\sim 10^{-2}$ (see Fig. 3 of Ref.~\cite{Carson:1990jm}) where it is nearly maximal, while it falls back again to $\sim 10^{-4}$ for $\lambda/g^2\sim 1$. Taking $\kappa$ close to its maximal value leads to the most restrictive requirement on $\bar{v}(T_C)/{T_C}$, raising the RHS of Eq.~(\ref{eq:washout3}) by $\sim 0.5$.
\item[(v)] The duration of the transition is often taken to be of order $t_H$, and we have correspondingly used $\Delta t_\mathrm{EW}/t_H=1$ as our benchmark value. However, the analysis of Ref.~\cite{Moore:1998swa} suggests that the transition could be as much as three orders of magnitude shorter. In this case, the requirement on ${\bar{v}(T_C)}/{T_C}$ could be weakened by nearly a factor of two.
\end{itemize}

In short, there exists considerable latitude in the requirement for baryon number preservation. Based on the foregoing numerical exploration we take the corresponding RHS of Eq.~~(\ref{eq:washout3}) to vary between 0.4 and 1.4 as a reasonable range for the Standard Model. This range depends on the value of the scalar self-coupling(s), the duration of the transition, and the degree of baryon number preservation needed as dictated by the CP-violating transport dynamics. In any realistic analysis, one must treat all of these considerations self-consistently.

\section{Discussion and Conclusions}
\label{sec:conclusions}
In this paper, we have identified the origin of gauge dependence of the BNPC.  We resolve these issues by (1) satisfying degeneracy and minimization conditions at each order in the loop expansion and (2) identifying the gauge-independent sphaleron scale which arises from dimensional reduction.  As an added bonus, our method to determine the critical temperature is substantially faster than the currently employed gauge dependent methods.

After a comparison of the gauge-independent $T_C$ in the SM with the results of lattice computations \cite{Csikor:1998ew}\cite{Moore:1998swa}, we find that the one-loop result appears to underestimate the value of the critical temperature and that a two-loop determination is likely to be required for a realistic perturbative approximation. Using existing results in the literature for $\veff(\phi,T)$ at two-loop order in the high temperature regime, we estimate the numerical impact of going beyond one-loop order. We emphasize, however, that a robust, gauge-independent treatment of $T_C$ at this order will require a future complete two-loop computation. When analyzing the dependence of the gauge-independent, perturbative computations of $T_C$ and ${\bar v}(T_C)$ on the underlying parameters, we also find that the results reproduce the trends obtained from Monte Carlo studies.

Nevertheless, we believe the precise quantitative implications of a perturbative computation for baryon number preservation are subject to considerable uncertainties, particularly for $T_C$ in the vicinity of $T_0$. Consequently, perturbative computations---while necessary when carrying out phenomenological investigations of the EWPT---should be taken as indicative rather than definitive. This need for such caution is even more apparent when one applies the results to the BNPC. 

Looking to future phenomenological studies that rely on perturbation theory, it is apparent that previous work relying on the use of Eq.~(\ref{eq:exp}) to determine the viability of EWB in given BSM scenarios should be revisited, taking into account the foregoing considerations: an appropriately gauge-independent determination of $T_C$; use of a gauge-independent scale---analogous to ${\bar v}(T)$---for the sphaleron rate; utilization of a appropriate values of $X$ and $\mathcal{F}$ in (\ref{eq:washout2}); and evaluation of the dependence of the fluctuation determinant $\kappa$ on the relevant couplings. As emphasized above, obtaining a realistic value of $T_C$ is likely to require computation of the two-loop, finite-temperature effective potential, at least in cases such as the MSSM wherein the dynamics of symmetry-breaking are dominated by loop effects. Determining the analog of ${\bar v}(T)$ will necessitate obtaining the sphaleron solution to the classical equations of motion in the presence of the extended scalar sector. Arriving at the appropriate value of $X$ will rely on the results of quantum transport computations that yield the initial baryon asymmetry, while obtaining the value of $\mathcal{F}$ will require more carefully identifying the degrees of freedom that dominate the unstable mode of fluctuations about the sphaleron. Together with the computation of $\kappa$, this program is likely to be numerically intensive. We expect that the results will at least be indicative of the regions of parameter space that are most likely to preserve an initial baryon asymmetry, pointing the way to more focused and robust non-perturbative calculations.

\begin{acknowledgements}
The authors thank S. Huber, A. J. Long and  M. Shaposhnikov for valuable discussions and helpful comments.
This work was supported in part by the U.S. Department of Energy contract DE-FG02-08ER41531 and by the Wisconsin Alumni Research Foundation.  MJRM also thanks the Aspen Center for Physics where a portion of this work was completed.
\end{acknowledgements}

\appendix
\section{Approximate BNPC}
\label{sec:washout}
In this appendix, we summarize the steps to derive the washout criterion in Eqn. (\ref{eq:exp}) in the standard model.  We are generally following the discussion found in the pedagogical review by Quir\'{o}s \cite{Quiros:1999jp}.

The SM expression of the baryon density $n_B$ depletion rate (\ref{eq:ratelaw}) is given by
\begin{equation}
\frac{d n_B}{d t}=-\frac{13n_f}{2}\frac{\Gamma_\text{sph}}{V T^3}\,n_B\,,
\end{equation}
where $n_f=3$ is the number of standard model generations.  Upon integrating, we find that the dilution at time $\Delta t_\text{EW}$ after the onset of the transition at time $t=0$ is
\begin{equation}\label{eq:dilute}
\frac{n_B(\Delta t_\text{EW})}{n_B(0)}=\exp\left[-\frac{13n_f}{2}\int_0^{\Delta t_\text{EW}}dt\,\frac{\Gamma_\text{sph}(T(t))}{V T^3(t)}\right]\,,
\end{equation}
where $T(t=0)=T_N\approx T_C$.  The washout criterion corresponds to imposing a lower bound $e^{-X}$ on the dilution factor
\begin{equation}
\frac{n_B(\Delta t_\text{EW})}{n_B(0)}>e^{-X}\,.
\end{equation}
To obtain an approximate expression for the washout criterion we assume that the integrand in (\ref{eq:dilute}) is approximately constant over the time of the transition.  After taking the double logarithm of both sides and using (\ref{eq:sphRate}) we find the following bound on $\bar{v}(T_C)/T_C$ (the baryon number preservation criterion --- BNPC):
\begin{multline}\label{eq:washout}
\frac{4\pi B}{g}\frac{\bar{v}(T_C)}{T_C}-6\ln\frac{\bar{v}(T_C)}{T_C} >\\
 -\ln X-\ln\Big(\frac{\Delta t_\text{EW}}{t_H}\Big)+\ln \mathcal{Z}+\hbar\ln\kappa\,,
\end{multline}
where we have chosen to normalize the duration of the phase transition $\Delta t_\text{EW}$ against the Hubble time $t_H$, and
\be
\mathcal{Z}=\left(\frac{13 n_f}{2}\right) \mathcal{N}_\mathrm{tr} \left(\mathcal{NV}\right)_\mathrm{rot}\left(\frac{\omega_{-}t_H}{\pi}\right)
\ee
contains the normalizations of the transitional and rotational modes of the sphaleron.  As emphasized in the main text, the ratio $\bar v(T_C)/T_C$ is gauge-independent only when the spahleron scale $\bar{v}(T)$ arising from dimensional reduction is used (section \ref{sec:sphaleron}), and when the critical temperature is computed in a manner consistent with Nielsen's identity (section \ref {GIanalysisSection}).

\section{Identities in the general model}\label{app:VGM}
In this appendix, we derive two useful theorems involving the field-dependent mass matrices in the general model, introduced in section \ref{SectionZeroT}.  The definitions of the mass matrices are reproduced here for convenience:
\begin{gather}
\begin{aligned}
M_{ij}^2(\phi)&=\frac{\partial^2 V(\phi)}{\partial\phi_i\partial\phi_j}\\
m_A^2(\phi)_{ij}&=(gT^a\phi)_i(gT^a\phi)_j\\
m_A^2(\phi)^{ab}&=(gT^a\phi)_i(gT^b\phi)_i,
\end{aligned}
\end{gather}
where $T^a$ are real-valued, and antisymmetric.
\begin{theorem}
Mass matrices $M_{ij}^2(\phi)$ and $m_A^2(\phi)_{ij}$ are simultanesouly diagonalizable when fields are set equal to their tree-level minimum.
\end{theorem}

\begin{proof}
Since $V(\Phi)$ belongs in the Lagrangian, it is invariant under transformations of $\mathcal{G}$,
\begin{equation}
V'(\Phi')=V(\Phi)\,,
\end{equation}
where $\Phi_i'=(1+\alpha^a gT^a)_{ij}\Phi_j$ is the infinitesimal transformation law for scalar fields. Then,
\begin{align}
\nonumber V'(\Phi')&=V'\big((1+\alpha^a gT^a)\Phi\big)\\
&= V(\Phi)+\alpha^a(gT^a\Phi)_i\frac{\partial V}{\partial \Phi_i}
\end{align}
implies
\begin{equation}
(gT^a\Phi)_i\frac{\partial V}{\partial \Phi_i}=0\,.
\end{equation}
We differentiate with respect to $\Phi_j$ to find \cite{Weinberg:1973ua}
\begin{equation}\label{massmtxiden}
gT^a_{ij}\frac{\partial V}{\partial\Phi_i}+(g T^a\Phi)_i M_{ij}^2(\Phi)=0\,,
\end{equation}
where we have used $\frac{\partial^2 V}{\partial\Phi_i\partial \Phi_j}=M^2_{ij}(\Phi)$.  When $\Phi_i$ is set equal to the tree-level minimum, $(\phi_0)_i$, the first term vanishes.  After multiplying from the left by $(gT^a\Phi)_k$ and summing over $a$ we find, after using $(gT^a\phi)_k(gT^a\phi)_i\equiv m_A^2(\phi)_{ki}$\,,
\begin{gather*}
m_A^2(\phi_0)_{ki} M_{ij}^2(\phi_0)=0\,,
\end{gather*}
that the mass matrices are not only simultaneously diagonalizable, but their eigenvalues live in distinct subspaces.
\end{proof}

\begin{theorem}
The gauge boson mass matrix $m_A^2(\phi)^{ab}$ and the gauge-fixing scalar boson mass matrix $m_A^2(\phi)_{ij}$ share identical non-zero eigenvalues.  This holds for any value of $\phi$.
\end{theorem}

\begin{proof}  
Let $\omega^b$ be an eigenvector of $m_A^2(\phi)^{ab}$ with eigenvalue $\lambda$:
\begin{equation}
m_A^2(\phi)^{ab}\omega^b\equiv (gT^a\phi)_i(g T^b\phi)_i\omega^b=\lambda\omega^a\,.
\end{equation}
Now multiply at left by $(gT^a\phi)_j$ and sum over $a$.
\begin{equation}
\underbrace{(gT^a\phi)_j(gT^a\phi)_i}_{m_A^2(\phi)_{ji}}(gT^b\phi)_i\omega^b=(gT^a\phi)_j\omega^a
\end{equation}
Then, defining $(gT^b\phi)_i\omega^b\equiv\mathcal{N}\omega_i$, we have
\begin{equation}
m_A^2(\phi)_{ji}\,\mathcal{N}\omega_i=\lambda \,\mathcal{N}\omega_j\,,
\end{equation}
that is, $\mathcal{N}\omega_i$ is an eigenvector of $m_A^2(\phi)_{ij}$ with the same eigenvalue $\lambda$.

Similarly, we could have started by letting $\omega_i$ be an eigenvector of $m_A^2(\phi)_{ij}$ with eigenvalue $\lambda$.  Then, by multiplying at left $(gT^b\phi)_i$, we find after defining $(gT^a\phi)_j\omega_j\equiv\mathcal{N}\omega^a$
\begin{equation}
m_A^2(\phi)^{ba}\,\mathcal{N}\omega^a=\lambda \,\mathcal{N}\omega^b\,,
\end{equation}
the same result.
\end{proof}

We note that although $(m_A^2)^{ab}$ and $(m_A^2)_{ij}$ have the same non-zero eigenvalues, multiplicities of the zero eigenvalues are expected to be different, as they have different dimensions.

\begin{widetext}
\section{High-$T$ expansion of $V_\text{eff}$ in the general model}\label{app:highTexp}
In this appendix, we derive the high-tempreature expansion of the finite temperature effective potential in the general model and prove that the $\mathcal{O}(T^2)$ terms (thermal masses) are gauge-independent.  We start with the expression given in (\ref{finiteTeffPot}),
\begin{equation}
V_\text{eff}(\phi,T)=V_\text{tree}(\phi)+V_\text{CW}(\phi)+\frac{T^4}{2\pi^2}\Big[\sum_{\text{scalar},i}\!\!J_B\left(m_i^2(\phi;\xi)/T^2\right)+3\sum_{\text{gauge},a}\!\!J_B\left(m_a^2(\phi)/T^2\right)-\sum_{\text{gauge},a}\!\!J_B\left(\xi m_a^2(\phi)/T^2\right)\Big]\,,
\end{equation}
and use the high temperature expansion of the thermal bosonic functions given in (\ref{eq:highTbosonic})
\begin{equation}
J_B(z^2)=-\frac{\pi^4}{45}+\frac{\pi^2}{12}z^2-\frac{\pi}{6}(z^2)^{3/2}-\frac{1}{32}z^4\ln z^2+\ldots\,,
\end{equation}
to obtain
\begin{multline}
V_\text{eff}(\phi,T)\approx V_\text{tree}(\phi)+V_\text{CW}(\phi)-\frac{\pi^2T^4}{90}(n_\text{s}+2n_\text{g})+\frac{T^2}{24}\Big[\sum_{\text{scalars,}i}\!\!m_i^2(\phi;\xi)+(3-\xi)\sum_{\text{gauge},a}\!\!m_a^2(\phi)\Big]\\
-\frac{T}{12\pi} \Big[\sum_{\text{scalars,}i}\!\!\big(m_i^2(\phi;\xi)\big)^{3/2}+(3-\xi^{3/2})\sum_{\text{gauge},a}\!\!\big(m_a^2(\phi)\big)^{3/2}\Big]
-\frac{1}{64\pi^2}\Big[\sum_{\text{scalars,}i}[m_i^2(\phi;\xi)]^2\ln\big(m_i^2(\phi;\xi)/T^2\big)\\
+3\sum_{\text{gauge},a}\big[m_a^2(\phi)\big]^2\ln\big(m_a^2(\phi)/T^2\big)-\sum_{\text{gauge},a}\big[\xi m_a^2(\phi)\big]^2\ln\big(\xi m_a^2(\phi)/T^2\big)\Big]\,,
\end{multline}
where $n_\text{s}$ and $n_\text{g}$ are numbers of scalar and gauge degrees of freedom as determined by the size of their respective mass matrices.  The last three terms proportional to $\ln(m^2/T^2)$ combine with the Coleman-Weinberg potential $V_\text{CW}(\phi)$ in (\ref{CWeffPot}) to cancel field-dependence inside the logarithms to generate terms of the form $\ln(T^2/\mu^2)$.  We chose the renormalization scale $\mu=T$ so that these logarithmic terms vanish.  The Stefan-Boltzmann $T^4$ terms are also independent of the mean field $\phi$. Thus, they do not influence the phase transition and may be dropped from the high-$T$ expansion.
Thus, at high temperature, the effective potential is dominated by $V_\text{tree}(\phi)$ and by $\mathcal{O}(T^2)$ and $\mathcal{O}(T)$ terms.
\begin{multline}\label{eq:highTexpGen}
V_\text{eff}(\phi,T)\approx V_\text{tree}(\phi)+\frac{T^2}{24}\Big[\sum_{\text{scalars,}i}\!\!m_i^2(\phi;\xi)+(3-\xi)\sum_{\text{gauge},a}\!\!m_a^2(\phi)\Big]\\
-\frac{T}{12\pi} \Big[\sum_{\text{scalars,}i}\!\!\big(m_i^2(\phi;\xi)\big)^{3/2}+(3-\xi^{3/2})\sum_{\text{gauge},a}\!\!\big(m_a^2(\phi)\big)^{3/2}\Big]
\end{multline}
We demonstrate the gauge-independence of $\mathcal{O}(T^2)$ terms in the following way:  since the sums run over the eigenvalues of the mass matrices, we may write them as traces of corresponding mass matrices.
\begin{align*}
\mathcal{O}(T^2)&=\frac{T^2}{24}\Big[\underset{ij}{\Tr}\big(M_{ij}^2(\phi)+\xi m_A^2(\phi)_{ij}\big)+(3-\xi)\underset{ab}{\Tr}\, m_A^2(\phi)^{ab}\Big]\\
&=\frac{T^2}{24}\Big[\Tr M_{ij}^2(\phi)+3\Tr m_A^2(\phi)^{ab}\Big]
\end{align*}
In the second line, we have used theorem 2 to cancel the gauge dependent terms, thus completing our explicit proof for gauge-independence of thermal masses at $\mathcal{O}(\hbar)$.

Therefore, the high-$T$ expansion of $V_\text{eff}$ reads
\begin{multline}\label{eq:genModelhighT}
V_\text{eff}(\phi,T)\approx V_\text{tree}(\phi)+\frac{T^2}{24}\Big[\Tr M_{ij}^2(\phi)+3\Tr m_A^2(\phi)^{ab}\Big]-\frac{T}{12\pi} \Big[\sum_{\text{scalars,}i}\!\!\big(m_i^2(\phi;\xi)\big)^{3/2}+(3-\xi^{3/2})\sum_{\text{gauge},a}\!\!\big(m_a^2(\phi)\big)^{3/2}\Big]
\end{multline}
In the language of (\ref{highTexp}), the $D$-coefficient is derived from the gauge-independent $\mathcal{O}(T^2)$ term; $T_0$ is, in part, governed by the tree-level mass terms present in $V_\text{tree}(\phi)$; the $E$-coefficient comes from the non-analytic $\mathcal{O}(T)$ terms; and the $\bar\lambda$-coefficient also comes from $V_\text{tree}(\phi)$ and additional terms not displayed in (\ref{eq:genModelhighT}).
\end{widetext}

\section{Gauge independence of the critical temperature}\label{app:critTemp}
The critical temperature is defined when two points in the effective potential are degenerate (\ref{defDegen}) and stationary (\ref{defMin}).  In this appendix we show that it follows from Nielsen's identity that the critical temperature is gauge-independent.

Let $V_\text{eff}(\phi_i,T;\xi)$ denote the gauge-dependent finite-temperature effective potential.  Suppose the phase transition takes the system from the first phase $\phi_i^{(1)}$ to the second phase $\phi_i^{(2)}$ at the critical temperature $T_c$ defined by conditions (\ref{defDegen}) and (\ref{defMin}) reproduced here,
\begin{gather}
\label{appDefDegen} V_\text{eff}(\phi^{(1)}_i,T_c;\xi)-V_\text{eff}(\phi_i^{(2)},T_c;\xi)=0\\
\label{appDefMin} \frac{V_\text{eff}}{\partial\phi_i}\Big|_{\phi^{(1)}_i,T_c}=\frac{\partial V_\text{eff}}{\partial\phi_i}\Big|_{\phi^{(2)}_i,T_c}=0\,.
\end{gather}
For now, we assume that simultaneously inverting these equations will yield gauge dependent field-values and critical temperature, $\phi^{(\bullet)}_i\equiv\phi^{(\bullet)}_i(\xi)$, and $T_c\equiv T_c(\xi)$.  A total differential of (\ref{appDefDegen}) with respect to $\xi$ gives
\begin{multline}
\Big(\frac{V_\text{eff}}{\partial\phi_i}\frac{\partial\phi_i^{(1)}}{\partial\xi}+\frac{\partial V_\text{eff}}{\partial T}\frac{\partial T_c}{\partial\xi}+\frac{\partial V_\text{eff}}{\partial\xi}\Big)_{\phi^{(1)},T_c}\\
-\Big(\frac{V_\text{eff}}{\partial\phi_i}\frac{\partial\phi_i^{(2)}}{\partial\xi}+\frac{\partial V_\text{eff}}{\partial T}\frac{\partial T_c}{\partial\xi}+\frac{\partial V_\text{eff}}{\partial\xi}\Big)_{\phi^{(2)},T_c}=0
\end{multline}
By the stationarity condition (\ref{appDefMin}) and Nielsen's identity (\ref{nielsenPotential}) the first and third terms vanish in each set of brackets, leaving us with
\begin{equation}
\Big(\frac{\partial V_\text{eff}}{\partial T}\Big|_{\substack{\phi^{(1)}\\T_c\,\,\,}}-\frac{\partial V_\text{eff}}{\partial T}\Big|_{\substack{\phi^{(2)}\\T_c\,\,\,}}\Big)\frac{\partial T_c}{\partial \xi}=0\,.
\end{equation}
The difference inside the brackets is, in general, non vanishing.  Hence we must have 
\begin{equation}\label{Tcgaugeindep}
\frac{\partial T_c}{\partial \xi}=0,
\end{equation} 
the critical temperature is gauge independent.

\section{Gauge dependence of scalar minimizing field $\phi_\text{min}$}\label{app:scCond}
In this section, we show that the field $\phi_\text{min}$ that minimizes the effective potential at finite temperature is gauge dependent.  Start by differentiating (\ref{appDefMin}) with respect to $\xi$.
\begin{equation}\label{ddv}
\frac{\partial^2 V_\text{eff}}{\partial\phi_i\,\partial\phi_j}\Big|_{\substack{\phi^{(1)}\\T_c\,\,\,}}\frac{\partial\phi_j^\text{(1)}}{\partial\xi}+\frac{\partial^2 V_\text{eff}}{\partial\phi_i\,\partial\phi_j}\Big|_{\substack{\phi^{(1)}\\T_c\,\,\,}}\frac{\partial T_c}{\partial\xi}+\frac{\partial^2 V_\text{eff}}{\partial\xi \partial\phi_i}\Big|_{\substack{\phi^{(1)}\\T_c\,\,\,}}=0
\end{equation}
By (\ref{Tcgaugeindep}) the second term vanishes.  We can re-express the third term by differentiating Nielsen's identity (\ref{nielsenPotential}), and evaluating the result at $\phi^{(1)}$ and $T_c$:
\begin{equation}\label{dNdp}
\frac{\partial^2 V_\text{eff}}{\partial\xi \partial\phi_i}=-\frac{\partial C_j}{\partial\phi_i}\frac{\partial V_\text{eff}}{\partial\phi_j}-C_j\frac{\partial^2 V_\text{eff}}{\partial\phi_i \partial\phi_j},
\end{equation}
where the second term vanishes at the minimum $\phi^{(1)}$.  So, (\ref{ddv}) becomes
\begin{equation}
\Big(\frac{\partial\phi_j^{(1)}}{\partial\xi}-C_j(\phi^{(1)},\xi)\Big)\frac{\partial^2 V_\text{eff}}{\partial\phi_i\partial\phi_j}\Big|_{\substack{\phi^{(1)}\\T_c\,\,\,}}=0
\end{equation}
Then, provided the curvature $\frac{\partial^2 V_\text{eff}}{\partial\phi_i\partial\phi_j}$ is non-vanishing at the minimum, we have
\begin{equation}\label{dpdx}
\frac{\partial\phi^{(1)}_j}{\partial\xi}=C_j(\phi^{(1)},\xi)\,.
\end{equation}
that the minimizing field is inherently gauge-dependent.

\begin{widetext}
\section{One-loop effective potential in the Standard Model}\label{app:VeffSM}
In this appendix, we provide expressions for the one-loop effective potential, 
with the dependence on the gauge parameter $\xi$ shown explicitly needed for the discussion in section \ref{sec:SM}.  In terms of the field dependent masses in (\ref{eq:SMfieldDepmasses}), the temperature-independent and -dependent parts of the one-loop potential are
\begin{multline}\label{eq:SMTindepVeff}
V_1^{T=0}(h)=\frac{1}{4(4\pi)^2}(m_H^2)^2\big[\ln(\textstyle\frac{m_H^2}{\mu^2})-\frac{3}{2}\displaystyle\big]
+\frac{2\times1}{4(4\pi)^2}(m_G^2+\xi m_W^2)^2\big[\ln(\textstyle\frac{m_G^2+\xi m_W^2}{\mu^2})-\frac{3}{2}\displaystyle\big]
\\
+\frac{1}{4(4\pi)^2}(m_G^2+\xi m_Z^2)^2\big[\ln(\textstyle\frac{m_G^2+\xi m_Z^2}{\mu^2})-\frac{3}{2}\displaystyle\big]
+\frac{2\times3}{4(4\pi)^2}(m_W^2)^2\big[\ln(\textstyle\frac{m_W^2}{\mu^2})-\frac{5}{6}\displaystyle\big]
+\frac{3}{4(4\pi)^2}(m_Z^2)^2\big[\ln(\textstyle\frac{m_Z^2}{\mu^2})-\frac{5}{6}\displaystyle\big]
\\
-\frac{2\times1}{4(4\pi)^2}(\xi m_W^2)^2\big[\ln(\textstyle\frac{\xi m_W^2}{\mu^2})-\frac{3}{2}\displaystyle\big]
-\frac{1}{4(4\pi)^2}(\xi m_Z^2)^2\big[\ln(\textstyle\frac{\xi m_Z^2}{\mu^2})-\frac{3}{2}\displaystyle\big]-\text{``free''}\,,
\end{multline}
and
\begin{multline}\label{eq:SMTdepVeff}
V^{T\neq0}_1(h,T)=\frac{T^4}{2\pi^2}\bigg[J_B\Big(\frac{m_H^2}{T^2}\Big)+2\!\times\!J_B\Big(\frac{m_G^2+\xi m_W^2}{T^2}\Big)+J_B\Big(\frac{m_G^2+\xi m_Z^2}{T^2}\Big)\bigg]\\
+\frac{3T^4}{2\pi^2}\bigg[2\!\times\!J_B\Big(\frac{m_W^2}{T^2}\Big)+J_B\Big(\frac{m_Z^2}{T^4}\Big)+J_B\Big(\frac{m_\gamma^2}{T^4}\Big)\bigg]
\\
-\frac{T^4}{2\pi^2}\bigg[2\!\times\!J_B\Big(\frac{\xi m_W^2}{T^2}\Big)+J_B\Big(\frac{\xi m_Z^2}{T^2}\Big)+J_B\Big(\frac{\xi m_\gamma^2}{T^2}\Big)\bigg]-\text{``free''}\,,
\end{multline}
where ``free'' represents a free-field subtraction.

\end{widetext}

\end{document}